\documentclass[reprint, superscriptaddress, amsmath, amssymb, aps, showkeys, pra, floatfix, longbibliography]{revtex4-2}

\usepackage{graphicx}
\usepackage{dcolumn}
\usepackage{bm}
\usepackage{float}
\usepackage{siunitx}
\usepackage{multirow}
\usepackage[colorlinks=true, linkcolor=red, citecolor=blue]{hyperref}
\begin{document}
\title{Theory of modulation instability in Kerr Fabry-Perot resonators beyond the mean field limit}
\author{Zoheir Ziani}\email{ziani.zoheir@gmail.com}
\author{Thomas Bunel}
\affiliation{University of Lille, CNRS, UMR 8523-PhLAM Physique des Lasers, Atomes et Molécules, F-59000, Lille, France}
\author{Auro M.~Perego}
\affiliation{Aston Institute of Photonic Technologies, Aston University, Birmingham, B4 7ET, UK}
\author{Arnaud Mussot}
\author{Matteo Conforti}\email{matteo.conforti@univ-lille.fr}
\affiliation{University of Lille, CNRS, UMR 8523-PhLAM Physique des Lasers, Atomes et Molécules, F-59000, Lille, France}
\date{\today}

\begin{abstract}
    We analyse the nonlinear dynamics of Fabry-Perot cavities of arbitrary finesse filled by a dispersive Kerr medium, pumped by a continuous wave laser or a synchronous train of flat-top pulses.
    The combined action of feedback, group velocity dispersion and Kerr nonlinearity leads to temporal instability with respect to perturbations at specified frequencies.
    We characterize the generation of new spectral bands by deriving the exact dispersion relation, and we find approximate analytical expressions for the instabilities threshold and gain spectrum of modulation instability (MI).
    We show that, in contrast to ring-resonators, both the stationary solutions and the gain spectrum are dramatically affected by the duration of the pump pulse.
    We derive the extended Lugiato-Lefever equation for the Fabry-Perot resonator (FP-LLE) starting from coupled nonlinear Schr\"odinger equations (rather than Maxwell-Bloch equations), and we compare the outcome of the stability analysis of the two models.
    While FP-LLE gives overall good results, we show regimes that are not captured by the mean-field limit, namely the period-two modulation instability, which may appear in highly detuned or nonlinear regimes.
    We report numerical simulations of the generation of MI-induced Kerr combs by solving FP-LLE and the coupled Schr\"odinger equations.
\end{abstract}
\keywords{Nonlinear optics, Resonator, Modulation instability, Linear stability analysis, Frequency combs}
\maketitle

\section{Introduction}
Optical cavities have been a valuable tool for studying various nonlinear effects since the invention of lasers in the 1960s.
Bistability, self-pulsing, and modulation instabilities are some examples of these effects that have been observed experimentally and analyzed theoretically \cite{lugiato2015nonlinear}.
Most of the early theoretical studies were focused on ring cavities, where the light propagates only in one direction, simplifying considerably the analysis \cite{ikeda1979multiple}.

Nonetheless, Fabry-Perot (FP) cavities, where two distinct fields propagate simultaneously in the forward and backward directions, are exploited in many applications.
Nonlinear interaction of counterpropagating fields can lead to very complex dynamics, even in the absence of a cavity.
For instance, it has been demonstrated that counterpropagation and nonlinearity can cause transverse spatial~\cite{firth1988transverse1,firth1990transverse2,firth1990transverse3,firth1992diffractive,geddes1994hexagons} and temporal instabilities~\cite{law1989dispersion,law1991instabilities}.
In resonators, temporal instabilities may appear even in the absence of group velocity dispersion (GVD), and they were first studied in a ring cavity (the well known Ikeda instability)~\cite{ikeda1979multiple} and later in FP systems~\cite{firth1981stability,abraham1982self,silberberg1984optical}.
Despite several attempts, dispersive instabilities (or temporal MI) in FP cavities are not completely characterized yet~\cite{yu1998temporal1,yu1998temporal2,firth2021analytic}.
A complete theoretical analysis has been developed only in the good-cavity (also called mean-field) approximation.
A version of the Lugiato-Lefever equation generalised to FP resonators (FP-LLE) has been derived, which permits to identify  the peculiarity of the FP case in an additional detuning term depending on the average field power \cite{cole2018theory}.
Beyond the mean field limit, analytical expressions of the MI threshold have been obtained for a specific resonator where one of the mirror has reflectivity equal to one \cite{firth2021analytic}.

The pioneering work on optical frequency combs (OFC) by Braje \textit{et al}.~\cite{braje2009brillouin} and subsequent research by Obrzud \textit{et al}.~\cite{obrzud2017temporal} in fiber-based FP cavities have opened up a new field of research focused on the generation and manipulation of OFC~\cite{bunel2023observation,jia2020photonic,xiao2023near,nie2022synthesized,xiao2020deterministic,wildi2023dissipative}.
They offer a high degree of flexibility in terms of comb bandwidth and mode spacing.
However, despite the advantages of using FP cavities to produce OFC, the experimental results are often poorly understood due to the lack of available analytical treatments.
The ongoing efforts in this field aim at improving the understanding of the physics of OFC and pave the way for their broader use in a variety of applications.

The goal of this paper is to describe MI in a nonlinear FP cavity with an instantaneous Kerr nonlinearity, second order GVD, arbitrary mirror reflectivity and arbitrary detuning.
We derive the full complex dispersion relation for the perturbations, which permits to calculate the \emph{exact} MI gain spectrum. We provide simpler but extremely accurate formulas of the MI gain, which extend the recent results reported in \cite{firth2021analytic}.
We compare the outcomes of our analysis with the prediction of the mean-field approximation, and report numerical simulations of MI comb generation in fiber FP resonators.
The paper is organised as follows.
In Sec.~\ref{sec:model}, we review the basic equations that describe FP cavities and derive the expression for the stationary solutions.
Then, in Sec.~\ref{sec:LSA}, we perform a linear stability analysis and obtain the exact dispersion relation.
Using appropriate approximations, we determine the gain spectrum that characterizes the modulation instability of homogeneous solutions.
In Sec.~\ref{sec:pulses}, we study the effect of pulsed pump on the system response.
Finally, in Sec.~\ref{sec:results}, we show some typical examples where mean-field approximation breaks down.
Conclusions are drawn in Sec.~\ref{sec:conclusion}.

\section{Fabry-Perot cavity description}\label{sec:model}

\begin{figure}[t]
    \centering
    \includegraphics[width=8.6cm]{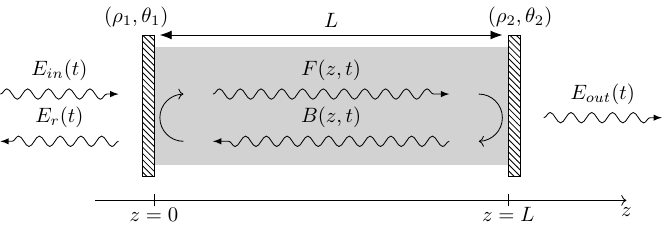}
    \caption{Schematic diagram of a nonlinear FP cavity of arbitrary finesse located in the interval $0 \leq z \leq L$.}
    \label{fig:FP_cavity}
\end{figure}

We consider a FP cavity of length $L$, filled with a nonlinear Kerr medium (see Fig.~\ref{fig:FP_cavity}).
A pump field $E_{in}$ enters at $z = 0$ through a mirror of reflectivity $\rho_1$ and drives forward $F(z, t)$ and backward $B(z, t)$ fields in the cavity.
A transmitted field $E_{out}$ exits the cavity through the second mirror of reflectivity $\rho_2$ at $z = L$.
The evolution of the two counterpropagating waves is described by a set of two coupled nonlinear Schrodinger equations~\cite{firth1981stability,firth2021analytic}~:
\begin{subequations}\label{eq:FP_2NLS}
    \begin{eqnarray}
        \frac{\partial F}{\partial z} + \beta_1 \frac{\partial F}{\partial t} + i\frac{\beta_2}{2}\frac{\partial^2 F}{\partial t^2} &=& i\gamma ( |F|^2 + G |B|^2)F\;,\label{eq:FP_2NLS_a}\\
        -\frac{\partial B}{\partial z} + \beta_1 \frac{\partial B}{\partial t} + i\frac{\beta_2}{2}\frac{\partial^2 B}{\partial t^2} &=& i\gamma ( |B|^2 + G |F|^2)B. \label{eq:FP_2NLS_b}
    \end{eqnarray}
\end{subequations}
where $\beta_1^{-1} = v_g$, is the group velocity, $\beta_2$ is the group-velocity dispersion coefficient, $\gamma$ is the nonlinear parameter and $G=2$ is the grating-parameter which describe cross-phase modulation (XPM).
The governing equations are supplemented with appropriate boundary conditions at the left and right mirrors~:
\begin{subequations}\label{eq:FP_BC}
    \begin{eqnarray}
        F(0,t) &=& \theta_1 E_{in}(t) + \rho_1 B(0,t)\;,\label{eq:FP_BC_a}\\
        B(L,t) &=& \rho_2 e^{i\phi_0} F(L,t).\label{eq:FP_BC_b}
    \end{eqnarray}
\end{subequations}
where the linear cavity phase $\phi_0$ account for the phase acquired during the propagation $2\beta_0L$ ($\beta_0$ is the propagation constant) and any possible contribution from the mirrors, modulo $2\pi$.
Thus $-\pi\leq\phi_0\leq\pi$, and we can introduce the cavity detuning as $\delta=-\phi_0$.
The transmitted field $E_{out}$ may be expressed as~:
\begin{equation}
    E_{out}(t) =  \theta_2 F(L,t).
\end{equation}
Thereafter, we assume that the reflectivity and the transmissivity of the mirrors are real and verify $\theta_{1,2}^2 + \rho_{1,2}^2 = 1$.
By taking $G=0$ and $\rho_2=1$, Eqs.~(\ref{eq:FP_2NLS}) and (\ref{eq:FP_BC}) model a ring-cavity of length $2L$.

Equations (\ref{eq:FP_2NLS_a}-\ref{eq:FP_BC_b}) have continuous wave (time-independent) solutions which are obtained by setting the time derivatives in (\ref{eq:FP_2NLS}) equal to zero and $E_{in}(t)$ constant ~\cite{firth1981stability,ogusu1998analysis,yu1998temporal2,firth2021analytic}.
They are of the form~:
\begin{subequations}
    \begin{eqnarray}
        F(z) &=& F_0 e^{+i\gamma (|F_0|^2 + G|B_0|^2)z} \equiv F_0 e^{i\phi_F z}\;, \\
        B(z) &=& B_0 e^{-i\gamma (|B_0|^2 + G|F_0|^2)z} \equiv B_0 e^{i\phi_B z}.
    \end{eqnarray}
\end{subequations}
Using (\ref{eq:FP_BC}) we find~:
\begin{subequations}\label{eq:FP_SS}
    \begin{eqnarray}
        F_0 &=& \frac{\theta_1 E_{in}}{1-\rho_1\rho_2 \exp{[i(\phi_0+\phi_{NL})}]}\;,\\
        B_0 &=& \rho_2 \exp{[i(\phi_0+\phi_{NL})]}F_0.
    \end{eqnarray}
\end{subequations}
where the nonlinear phase is given by~:
\begin{equation}
    \phi_{NL} = \gamma (1+\rho_2^2)(1+G)L |F_0|^2.
\end{equation}
From Eqs.~(\ref{eq:FP_SS}) we obtain the input power $P_{in}=|E_{in}|^2$ as a function of the intracavity forward power $P_F = |F_0|^2$~:
\begin{equation}\label{eq:SS_power}
    P_{in} = \frac{P_F}{\theta_1^2} \left( 1 + (\rho_1\rho_2)^2 -2\rho_1\rho_2 \cos(\phi_0 + \phi_{NL})\right).
\end{equation}
From Eq.~(\ref{eq:SS_power}) we find that the cavity finesse, i.e. the ratio between the line-width and the free spectral range (FSR) is given by  $\mathcal{F}=\dfrac{\pi\sqrt{\rho_1\rho_2}}{1-\rho_1\rho_2}$.

\section{Linear stability analysis}\label{sec:LSA}

\subsection{General dispersion relation}
Stability of the steady state is examined assuming a time-dependent solution of the form~:
\begin{subequations}
    \begin{eqnarray}
        F(z,t) &=& F_0 (1 + f(z,t)) e^{i\phi_F z}\;,\\
        B(z,t) &=& B_0 (1 + b(z,t)) e^{i\phi_B z}\;,
    \end{eqnarray}
\end{subequations}
where $f$ and $b$ are small perturbations.
The linearized propagation equations for the perturbations read as~:
\begin{subequations}\label{eq:linearized_eqs}
\begin{eqnarray}
    \frac{\partial f}{\partial z} + \beta_1 \frac{\partial f}{\partial t} + i\frac{\beta_2}{2}\frac{\partial^2 f}{\partial t^2} &=&2i\gamma P_F \Re\left( f + \rho_2^2 G b\right),\\
    -\frac{\partial b}{\partial z} + \beta_1 \frac{\partial b}{\partial t} + i\frac{\beta_2}{2}\frac{\partial^2 b}{\partial t^2} &=&2i\gamma P_F \Re\left( G f + \rho_2^2 b \right).
\end{eqnarray}
\end{subequations}
$\Re$ stands for the real part.
From Eqs.~(\ref{eq:FP_BC}), we find the following boundary conditions for $f$ and $b$~:
\begin{subequations}
    \begin{eqnarray}
        f(0,t) &=& \rho_1\rho_2 e^{i\phi} b(0,t)\;,\\
        f(L,t) &=& b(L,t),
    \end{eqnarray}
\end{subequations}
where $\phi = \phi_0+\phi_{NL}$.
We write the perturbation in the following form~:
\begin{subequations}\label{eq:perturbation_forms}
    \begin{eqnarray}
        f(z,t) &=& f_+(z) e^{\lambda t} + f_-^*(z) e^{\lambda^* t}\;, \\
        b(z,t) &=& b_+(z) e^{\lambda t} + b_-^*(z) e^{\lambda^* t}.
    \end{eqnarray}
\end{subequations}
The real and imaginary parts of $\lambda=\sigma+i\omega$ defines the temporal growth rate and the frequency of the perturbations.
By inserting Eqs.~(\ref{eq:perturbation_forms}) in Eqs.~(\ref{eq:linearized_eqs}), we find that the four complex amplitudes $f_\pm$, $b_\pm$ of the perturbations obey the following system of ordinary differential equations~:
\begin{equation}\label{eq:MED}
    \frac{d}{d z}
    \begin{pmatrix}
        f_+(z) \\ f_-(z) \\b_+(z) \\b_-(z)
    \end{pmatrix}
    =
    \mathcal{M}
    \begin{pmatrix}
        f_+(z) \\ f_-(z) \\b_+(z) \\b_-(z)
    \end{pmatrix}
\end{equation}
where
\begin{align*}
&\mathcal{M} = \\
&i\gamma P_F
\begin{pmatrix}
    1+i\frac{\psi_+}{\gamma P_F} & 1 & G\rho_2^2 & G\rho_2^2 \\
    -1 & -1+i\frac{\psi_-}{\gamma P_F} & -G \rho_2^2 & -G\rho_2^2 \\
    -G & -G & -\rho_2^2-i\frac{\psi_+}{\gamma P_F} & -\rho_2^2 \\
    G & G & \rho_2^2 & \rho_2^2-i\frac{\psi_-}{\gamma P_F}
\end{pmatrix} ,
\end{align*}
with $\psi_\pm = \beta_1\lambda \pm i\dfrac{\beta_2}{2}\lambda^2$.
The growth rate $\sigma$ and the frequency $\omega$ of the perturbations are found by imposing the following boundary conditions~:
\begin{subequations}\label{eq:perturbation_forms2}
    \begin{eqnarray}
        f_\pm(L) &=& b_\pm(L)\;,\\
        f_\pm(0) &=& \rho_1\rho_2 e^{\pm i\phi} b_\pm(0).
    \end{eqnarray}
\end{subequations}
The matrix differential equation~(\ref{eq:MED}) is linear and homogeneous, so it can be solved by standard methods (e.g. matrix exponential).
However, the analytic expressions are very cumbersome since they depends on the roots of a general 4th order polynomial. 
The eigenvalues $\eta_j$ ($j=1-4$) of $\mathcal{M}$ are given by the roots of the characteristic polynomial~:
\begin{equation}
    \eta_j^4 + a_2 \eta_j^2 + a_1 \eta_j + a_0 = 0
\end{equation}
where
\begin{eqnarray*}
    a_2 &=& \frac{\beta_2^2}{2}\lambda^4-2\beta_1^2\lambda^2-\gamma P_F (1+\rho_2^2)\beta_2\lambda^2\;,\\
    a_1 &=& 2\gamma P_F \beta_1\beta_2 (1-\rho_2^2)\lambda^3\;\\
    a_0 &=& \frac{\beta_2^4}{16}\lambda^8 + \frac{\beta_2^2}{4} (2\beta_1^2-\gamma P_F (1+\rho_2^2)\beta_2 )\lambda^6 \\
    & & + (\beta_1^4 -\gamma P_F \beta_1^2\beta_2 (1+\rho_2^2) -\gamma^2 P_F^2 \rho_2^2 (G^2-1)\beta_2^2)\lambda^4.
\end{eqnarray*}
The solution of Eqs.~(\ref{eq:MED}) can be expressed in terms of eigenvalues and eigenvectors of $\mathcal{M}$~:
\begin{equation}\label{eq:LSA_solution}
        (f_+, f_-, b_+, b_-)^T  = \sum_{j=1}^4 c_j e^{\eta_j z} \textbf{u}_j,
\end{equation}
where $\textbf{u}_j =(u_{j,1},u_{j,2},u_{j,3},u_{j,4})^T$ are the eigenvectors of $\mathcal{M}$.
The four arbitrary constants $c_j$ are determined by the boundary conditions.
\begin{widetext}
    \noindent Indeed, by inserting Eq.~(\ref{eq:LSA_solution}) in Eq.~(\ref{eq:perturbation_forms2}), we obtain a system of algebraic equations $\mathcal{N} (c_1,c_2,c_3,c_4)^T = 0$, where
    \begin{equation}
        \mathcal{N} = 
        \begin{pmatrix}
            (u_{1,1}-u_{1,3})e^{\eta_1 L} &  (u_{2,1}-u_{2,3})e^{\eta_2 L} & (u_{3,1}-u_{3,3})e^{\eta_3 L} & (u_{4,1}-u_{4,3})e^{\eta_4 L} \\
            (u_{1,2}-u_{1,4})e^{\eta_1 L} &  (u_{2,2}-u_{2,4})e^{\eta_2 L} & (u_{3,2}-u_{3,4})e^{\eta_3 L} & (u_{4,2}-u_{4,4})e^{\eta_4 L} \\
            u_{1,1}-\rho_1\rho_2 e^{+i\phi} u_{1,3} & u_{2,1}-\rho_1\rho_2 e^{+i\phi} u_{2,3} & u_{3,1}-\rho_1\rho_2 e^{+i\phi} u_{3,3} & u_{4,1}-\rho_1\rho_2 e^{+i\phi} u_{4,3} \\
            u_{1,2}-\rho_1\rho_2 e^{+i\phi} u_{1,4} & u_{2,2}-\rho_1\rho_2 e^{+i\phi} u_{2,4} & u_{3,2}-\rho_1\rho_2 e^{+i\phi} u_{3,4} & u_{4,2}-\rho_1\rho_2 e^{+i\phi} u_{4,4} 
        \end{pmatrix}.
    \end{equation}
    In order to have a nontrivial solution, we must impose determinant of $\mathcal{N}$ to be zero, which yields~:
    \begin{equation}\label{eq:FP_DR_general}
        \det \mathcal{N}= C_{1,2} e^{(\eta_1+\eta_2)L} + C_{1,3} e^{(\eta_1+\eta_3)L} + C_{1,4} e^{(\eta_1+\eta_4)L} + C_{2,3} e^{(\eta_2+\eta_3)L} + C_{2,4} e^{(\eta_2+\eta_4)L} + C_{3,4} e^{(\eta_3+\eta_4)L} = 0,
    \end{equation}
    where
    \begin{eqnarray*}
        C_{1,2} &=& +[ u_{3,1}u_{4,2} -u_{3,2}u_{4,1} +\rho_1^2\rho_2^2 (u_{3,3}u_{4,4}-u_{3,4}u_{4,3}) -\rho_1\rho_2 \left( (u_{3,1}u_{4,4}-u_{3,4}u_{4,1})e^{-i\phi} -(u_{3,2}u_{4,3}-u_{3,3}u_{4,2})e^{+i\phi}\right)] \sigma_{1,2}\;\\
        C_{1,3} &=& -[ u_{2,1}u_{4,2} -u_{2,2}u_{4,1} +\rho_1^2\rho_2^2 (u_{2,3}u_{4,4}-u_{2,4}u_{4,3}) -\rho_1\rho_2 \left( (u_{2,1}u_{4,4}-u_{2,4}u_{4,1})e^{-i\phi} -(u_{2,2}u_{4,3}-u_{2,3}u_{4,2})e^{+i\phi}\right)] \sigma_{1,3}\;\\
        C_{1,4} &=& +[ u_{2,1}u_{3,2} -u_{2,2}u_{3,1} +\rho_1^2\rho_2^2 (u_{2,3}u_{3,4}-u_{2,4}u_{3,3}) -\rho_1\rho_2 \left( (u_{2,1}u_{3,4}-u_{2,4}u_{3,1})e^{-i\phi} -(u_{2,2}u_{3,3}-u_{2,3}u_{3,2})e^{+i\phi}\right)] \sigma_{1,4}\;\\
        C_{2,3} &=& +[ u_{1,1}u_{4,2} -u_{1,2}u_{4,1} +\rho_1^2\rho_2^2 (u_{1,3}u_{4,4}-u_{1,4}u_{4,3}) -\rho_1\rho_2 \left( (u_{1,1}u_{4,4}-u_{1,4}u_{4,1})e^{-i\phi} -(u_{1,2}u_{4,3}-u_{1,3}u_{4,2})e^{+i\phi}\right)] \sigma_{2,3}\;\\
        C_{2,4} &=& -[ u_{1,1}u_{3,2} -u_{1,2}u_{3,1} +\rho_1^2\rho_2^2 (u_{1,3}u_{3,4}-u_{1,4}u_{3,3}) -\rho_1\rho_2 \left( (u_{1,1}u_{3,4}-u_{1,4}u_{3,1})e^{-i\phi} -(u_{1,2}u_{3,3}-u_{1,3}u_{3,2})e^{+i\phi}\right)] \sigma_{2,4}\;\\
        C_{3,4} &=& +[ u_{1,1}u_{2,2} -u_{1,2}u_{2,1} +\rho_1^2\rho_2^2 (u_{1,3}u_{2,4}-u_{1,4}u_{2,3}) -\rho_1\rho_2 \left( (u_{1,1}u_{2,4}-u_{1,4}u_{2,1})e^{-i\phi} -(u_{1,2}u_{2,3}-u_{1,3}u_{2,2})e^{+i\phi}\right)] \sigma_{3,4}\;\\
    \end{eqnarray*}
\end{widetext}
and
\begin{equation*}
    \sigma_{i,j} = (u_{i,1}-u_{i,3})(u_{j,2}-u_{j,4})-(u_{j,1}-u_{j,3})(u_{i,2}-u_{i,4}).
\end{equation*}
The dispersion relation given Eq.~(\ref{eq:FP_DR_general}) is a complicated nonlinear equation in the complex variable $\lambda =\sigma+i\omega$.
Even if we are not able to solve it analytically, it can be solved numerically \footnote{We found more convenient to calculate the eigenvectors of matrix $\mathcal{M}$ numerically. In order to avoid numerical instabilities, we have used the Matlab built-in function \texttt{eig} by explicitly choosing the algorithm used, i.e. \texttt{[V,D] = eig(M,eye(4),'chol')}}.
Its solutions in the complex $\lambda$-plane give the growth rate and the frequency of the possibly unstable perturbations.

\subsection{Approximate solutions}\label{sec:approx_sols}
The solution of Eqs.~(\ref{eq:MED}) is greatly simplified if $\beta_2 = 0$ or if $G = 0$.
For the dispersion-less case, we obtain the same results of Firth~\cite{firth1981stability} (calculation not reported here).
For the dispersive case, in order to achieve reasonably simple analytical expressions, we assume $G=0$ in Eqs.~(\ref{eq:MED}) only.
It amounts to suppressing the linear coupling between the perturbations components in the propagation equations.
Nonetheless, the coupling is maintained in the boundary conditions via the total phase $\phi$.
This approximation is physically sound~: the main coupling between forward and backward perturbations takes place at the mirrors.
We also maintain $G\neq 0$ in the steady state, meaning that the perturbations propagate on top of the correct steady state.
Following similar arguments, the same approximation has been used in \cite{yu1998temporal1,yu1998temporal2,firth2021analytic}.

Solving Eqs.~(\ref{eq:MED}) with $G=0$ and using the same approach as before, we find the following characteristic equation~:
\begin{equation}\label{eq:FP_approx}
    e^{2\lambda t_R} -\Delta(\lambda)e^{\lambda t_R} + (\rho_1\rho_2)^2 = 0\;,
\end{equation}
where $t_R =2\beta_1L$ is the roundtrip time and
\begin{eqnarray*}
    \Delta(\lambda) &=& \rho_1\rho_2 [a(\lambda)\cos\phi + b(\lambda)\sin\phi]
\end{eqnarray*}
with 
\begin{eqnarray*}
    a(\lambda) &=& 2\cos(kL)\cos(k_\rho L) - \frac{k^2+k_\rho^2}{k k_\rho} \sin(kL)\sin(k_\rho L)\;,\\
    b(\lambda) &=& \frac{4k^2+\beta_2^2\lambda^4}{2\beta_2\lambda^2k}\cos(k_\rho L)\sin(kL) \\ 
    & &\hskip0.1\textwidth+\frac{4k_\rho^2+\beta_2^2\lambda^4}{2\beta_2\lambda^2k_\rho}\cos(k L)\sin(k_\rho L),
\end{eqnarray*}
and
\begin{align}\label{eq:k_omega}
    \nonumber k^2 &= \frac{\beta_2\lambda^2}{2} \left(\frac{\beta_2\lambda^2}{2}-2\gamma P_F \right)\;, \\ 
    k_\rho^2 &= \frac{\beta_2\lambda^2}{2} \left(\frac{\beta_2\lambda^2}{2}-2\gamma P_F\rho_2^2 \right).
\end{align}
Even if the structure of Eq.~(\ref{eq:FP_approx}) appears to be quite simple, it may admit an infinite number of solutions, since the discriminant depends on $\lambda$.
We may also remark that Eq.~(\ref{eq:FP_approx}) is equivalent to the following two equations~:
\begin{equation}
    e^{\lambda t_R} = \frac{\Delta(\lambda)}{2} \pm \sqrt{\frac{\Delta^2(\lambda)}{4}-(\rho_1\rho_2)^2}.
\end{equation}
From the analysis of the dispersion-less case~\cite{firth1981stability}, we have learned that $\omega t_R = m \pi$ at threshold ($\sigma=0$).
It is worth note that the frequencies of the perturbations at threshold correspond either to cavity resonances ($m$ even) or they are in between ($m$ odd, anti-resonance).
For $\sigma=0$, Eq.~(\ref{eq:FP_approx}) reads as
\begin{equation}
    1 - (-1)^m \Delta + (\rho_1\rho_2)^2 =0,
\end{equation}
which gives the analytic threshold expression 
\begin{equation}\label{eq:MI_thresholds}
    \tilde{a}(\omega) \cos\phi + \tilde{b}(\omega)\sin\phi = (-1)^m \frac{1 + (\rho_1\rho_2)^2}{\rho_1\rho_2},
\end{equation}
where $\tilde{a}(\omega) = a(\lambda=i\omega)$ and $\tilde{b}(\omega) = b(\lambda=i\omega)$.
By taking $\rho_2 = 1$, we recover the main result of \cite{firth2021analytic} in which the authors used the gain-circle method to find the threshold formula for zero-transmission case.

\begin{figure}[t]
    \centering
    \includegraphics[width=8cm]{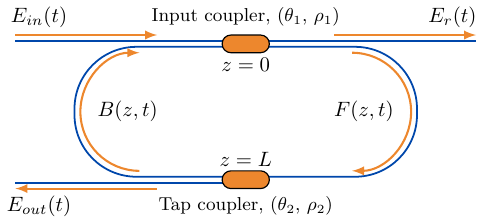}
    \caption{The Fabry-Perot cavity of Fig.~\ref{fig:FP_cavity} in the limit $G=0$ is equivalent to a ring cavity composed of two identical pieces of fibers of length $L$ connected by two couplers ($\theta_1$,$\rho_1$) and ($\theta_2$,$\rho_2$).}
    \label{fig:ring_resonator}
\end{figure}

Numerical solution of Eq.~(\ref{eq:FP_approx}), shows that $\omega t_R \approx m \pi$ approximately holds even when $\sigma\neq0$ with great precision.
Moreover, we may assume that the frequency of the perturbation is much greater than its growth rate, i.e. $\omega \gg \sigma$. We thus write Eq.~(\ref{eq:FP_approx}) in the form
\begin{equation}
    e^{2\sigma t_R} -(-1)^m \tilde\Delta(\omega)e^{\sigma t_R} + (\rho_1\rho_2)^2 = 0,
\end{equation}
with $\tilde\Delta(\omega)=\Delta(\lambda=0+i\omega)$.

It is now straightforward to calculate the growth rate of the perturbations as a function of the frequency,~i.e.~$\sigma(\omega)$
\begin{equation}
    e^{\sigma t_R} = (-1)^m \frac{\tilde\Delta(\omega)}{2} \pm \sqrt{\frac{\tilde\Delta^2(\omega)}{4}-(\rho_1\rho_2)^2}.
\end{equation}
In order to observe MI, we must have $\sigma>0$, so we have the following conditions
\begin{eqnarray}
(-1)^m \tilde\Delta(\omega) >1 +(\rho_1\rho_2)^2,
\end{eqnarray}
and
\begin{equation}\label{eq:str}
    \sigma t_R = \ln\left((-1)^m\frac{\tilde\Delta}{2}+ \sqrt{\frac{\tilde\Delta^2}{4}-(\rho_1\rho_2)^2} \right)
\end{equation}
We can define the MI gain $g(\omega)$ as the spatial growth rate~:
\begin{equation}\label{eq:FP_gain}
    g(\omega) = \frac{\sigma t_R}{2L} = \frac{1}{2L} \ln \max \left|\frac{\tilde\Delta}{2}\pm \sqrt{\frac{\tilde\Delta^2}{4}-(\rho_1\rho_2)^2}\right|
\end{equation}
We recognize in Eq.~(\ref{eq:FP_gain}) the MI gain of a ring cavity of length $2L$ composed of two identical pieces of fiber of length $L$~\cite{conforti2016parametric} connected by an input coupler ($\theta_1$,$\rho_1$), and a tap coupler ($\theta_2$,$\rho_2$) as illustrated in Fig.~\ref{fig:ring_resonator} (See Appendix~\ref{sec:ringMI} for details).

\begin{figure}[t]
    \centering
    \includegraphics[width=8.6cm]{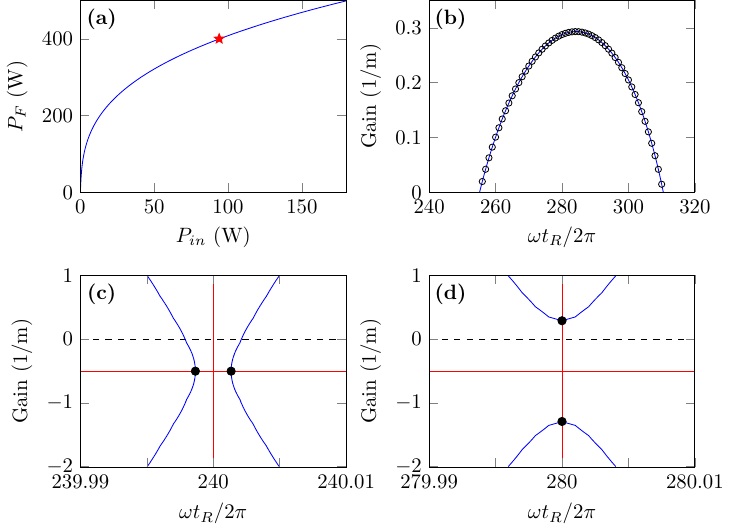}
    \caption{(a) Intracavity forward power as a function of pump power from Eq.~(\ref{eq:SS_power}).
            (b) Approximated MI gain $g(\omega)$ from Eq.~(\ref{eq:FP_gain}) (blue curve) and exact gain from numerical solution of Eq.~(\ref{eq:FP_DR_general}) (black circles) (c) Real (blue curve) and imaginary (red curve) part of Eq.~(\ref{eq:FP_DR_general}) outside the MI gain band.
            The intersections between the two curves (black circles) are the solutions of Eq.~(\ref{eq:FP_DR_general}).
            (d) Same as (c) but inside the MI gain band.
            Parameters: $\rho_1^2=\rho_2^2=0.99$ ($\mathcal{F}=312$), $\theta_1^2=\theta_2^2=0.01$, $\gamma = \SI{2}{W^{-1}/km}$, $\beta_1=c/1.5$, $\beta_2 = \SI{-20}{ps^2/km}$, $L= \SI{0.01}{m}$, $\phi_0=0$, $P_{in}=\SI{94}{W}$ and $P_{F}=\SI{400}{W}$.}
    \label{fig:exactMIgain}
\end{figure}

We verified that the approximation Eq.~(\ref{eq:FP_gain}) is extremely precise.
As an example, we report in Fig.~\ref{fig:exactMIgain} the comparison between the exact and the approximated MI gain for a fiber FP resonator with an anomalous dispersion fiber and operating in the monostable regime (parameters are reported in the figure's caption).
Figure~\ref{fig:exactMIgain}(a) shows the steady-state curve Eq.~(\ref{eq:SS_power}), the working point being denoted with a red star.
Figure~\ref{fig:exactMIgain}(b) shows the approximated MI gain from Eq.~(\ref{eq:FP_gain}) (blue curve) and the exact gain from numerical solution of Eq.~(\ref{eq:FP_DR_general}) (black circles).
The circles are perfectly superposed to the continuous line: there is no visible difference between the two models.
Indeed, we have found that for all our test the exact and the approximated model give essentially identical results.
Of course, the approximated model does not predict the frequencies of the perturbations, because here $\omega$ is a continuous, independent variable.

Two examples of the graphical solution of Eq.~(\ref{eq:FP_DR_general}) are shown in Fig.~\ref{fig:exactMIgain}(c,d).
Blue and red curves represent $\Re(\det \mathcal{N})=0$ and $\Im(\det \mathcal{N})=0$ in the complex $\lambda$-plane.
As a general feature, the curve $\Im(\det \mathcal{N})=0$ is essentially composed of an horizontal line with $\sigma<0$ and a set of vertical lines at $\omega t_R\approx m\pi$.
The solutions are marked by solid black dots.
Figure~\ref{fig:exactMIgain}(c) shows two solutions with $\sigma<0$, meaning that the perturbations at the corresponding frequency $\omega$ are stable.
Whereas, Fig.~\ref{fig:exactMIgain}(d) shows two solutions which share almost the same frequency, (they are slightly different and very close to a cavity resonance $\omega t_R=2m\pi$) but opposite $\sigma$.
The fact that one solution has $\sigma>0$ implies that the perturbation at this frequency is unstable.

\subsection{The good-cavity limit: FP-LLE} \label{sec:FPLLE}
When the mirror reflectivities are high and the cavity detuning is small, it is possible to obtain a mean-field description of the dynamics, which generalises the celebrated Lugiato-Lefever equation (LLE) originally derived for the ring resonators \cite{lugiato1987spatial,haelterman1992additive} to FP cavities.
Cole \textit{et~al}.~\cite{cole2018theory} derived this FP-LLE starting form Maxwell-Bloch equations, while Xiao \textit{et al}.~\cite{xiao2020deterministic} also arrived at the same equation from coupled mode theory.
In appendix~\ref{sec:LLE}, we report an alternative derivation of the FP-LLE, which uses the coupled NLS equations~(\ref{eq:FP_2NLS}) as the starting point.
Besides being more suited to fiber-based FP resonators, our derivation is more general as it considers unequal mirror reflectivities and pulsed pumping. 
The FP-LLE reads as~:
\begin{align}
  \nonumber  & t_R \frac{\partial \psi}{\partial \tau} = (-\alpha +i\phi_0)\psi + \theta_1 E_{in} \\
  &+ 2L \left[ -i\frac{\beta_2}{2}\frac{\partial^2}{\partial t^2} +i\gamma |\psi|^2 +i\gamma\frac{G}{t_R} \int_{-t_R/2}^{t_R/2}|\psi|^2dt\right] \psi, \label{eq:FP_LLE}
\end{align}
where $\psi(\tau,t)$ is the field envelope inside the cavity, $\alpha~=~1-\rho_1\rho_2$ is the cavity loss, $t \in [-t_R/2, t_R/2]$ denotes the fast time in one cavity roundtrip and $\tau$ is a slow time.

The homogeneous solutions $\psi_s$ are found by setting the derivatives in Eq.~(\ref{eq:FP_LLE}) equal to zero. 
We obtain that the power of the stationary solutions is given by the solutions of the cubic following equation~:
\begin{equation}\label{eq:LLE_SS}
    \theta_1^2 P_{in} = P_s \left( \alpha^2 + (\phi_0 + 2\gamma L (1+G)P_s )^2 \right),
\end{equation}
where $P_s=|\psi_s|^2$ and $P_{in}=|E_{in}|^2$.
We can assume $\psi_s$ real without loss of generality, which implies that the input field must be complex and can be written as~:
\begin{equation}
    \theta_1 E_{in} = \psi_s ( \alpha -i(\phi_0+2\gamma L (1+G)P_s ) ).
\end{equation}

To study the stability of these solutions, we perform a linear stability by considering a perturbed solution of the form $\psi(t,\tau) = \psi_s + \xi(\tau,t)$.
Assuming $\xi\ll\psi_s$ small, we obtain
\begin{align}
 \nonumber   & t_R \frac{\partial \xi}{\partial \tau} = (-\alpha+i\phi_0)\xi -iL\beta_2 \frac{\partial^2 \xi}{\partial t^2} + \\
 & 2i\gamma L P_s\left( (2+G)\xi + \xi^* + \frac{G}{t_R} \int_{-t_R/2}^{t_R/2}(\xi+\xi^*)dt\right).
\end{align}
We now expand the perturbation over the cavity modes with time-varying amplitudes~:
\begin{equation}
    \xi(\tau,t) = \varepsilon_n(\tau)e^{i\omega_n t} + \varepsilon_{-n}(\tau)e^{-i\omega_n t},
\end{equation}
with $\omega_n=n2\pi/t_R$.
The amplitudes of the modal perturbations obey
\begin{subequations}\label{eq:LLE_LS_LE}
    \begin{align}
       \nonumber & t_R\frac{d \varepsilon_n}{d\tau} = (-\alpha+i\phi_0)\varepsilon_n +iL\beta_2\omega_n^2 \varepsilon_n \\
        & +2i\gamma L P_s ( (2+G)\varepsilon_n + \varepsilon_{-n}^*) + 2i\gamma L P_s G (\varepsilon_n + \varepsilon_{-n}^*)\delta_{n0}\\
       \nonumber  & t_R\frac{d \varepsilon_{-n}^*}{d\tau} = (-\alpha-i\phi_0)\varepsilon_{-n}^* +iL\beta_2\omega_n^2 \varepsilon_{-n}^* \\
        &-2i\gamma L P_s ( (2+G)\varepsilon_{-n}^* + \varepsilon_n) - 2i\gamma L P_s G (\varepsilon_n + \varepsilon_{-n}^*)\delta_{n0}
    \end{align}
\end{subequations}
where $\delta_{n0}$ is the Kröneker delta.
The last terms in Eqs.~(\ref{eq:LLE_LS_LE}), which appears only for the zero mode, stem from the integral term which do not average zero as in the case $n\neq 0$.
This contribution is not present for the ring cavity, for which $G=0$.
The system~(\ref{eq:LLE_LS_LE}) can be written as $d/d\tau(\varepsilon_n, \varepsilon_{-n}^*)^T = M_n (\varepsilon_n, \varepsilon_{-n}^*)^T$, and the eigenvalues of the matrix $M_n$ determine the stability of the solution.
The temporal growth rate of the perturbations for $n\neq 0$ reads~:
\begin{equation}\label{eq:FP_LLE_gain}
    \sigma(\omega_n) = \frac{1}{t_R} (-\alpha + \sqrt{(2\gamma L P_s)^2 -\mu_n^2}),
\end{equation}
where $\mu_n = \phi_0 +L\beta_2\omega_n^2 + 2(2+G)\gamma L P_s$.
The temporal growth can be written as a spatial gain as $g(\omega_n) = \beta_1 \sigma(\omega_n)$.
The most unstable mode, obtained for $\mu_n =0$, and its growth rate are
\begin{equation}\label{eq:FP_LLE_gmax}
    \omega_{\Bar{n}}^2 =-\frac{\phi_0+2(2+G)\gamma L P_s}{\beta_2 L}\, \quad\text{and} \quad g_{max} =  \frac{-\alpha +2\gamma L P_s}{2L},
\end{equation}
where we considered $\omega_n$ as a continuous variable.
We can thus interpret the condition $\mu_n=0$ as a phase-matching relation that maximises the energy transfer from the pump to the perturbations.

For the zero mode we have
\begin{equation}\label{eq:LLE_mode0}
    \sigma(0) = \frac{-\alpha + \sqrt{(2\gamma L P_s (1+G))^2 -(\phi_0+4(1+G)\gamma L P_s)^2}}{t_R}.
\end{equation}
The condition for reality of Eq.~(\ref{eq:LLE_mode0}) coincides with the negative slope branch of Eq.~(\ref{eq:LLE_SS}), i.e. the homogeneous solution is unstable if $P^-<P_s<P^+$, where 
\begin{equation}\label{eq:LLE_BS}
    P^\pm = \frac{-2\phi_0 \pm \sqrt{\phi_0^2-3\alpha^2}}{6(1+G)\gamma L}.
\end{equation}
The unstable region obtained by letting $\omega_n \rightarrow 0$ in Eq.~(\ref{eq:FP_LLE_gain}) is different and the limits are given by 
\begin{equation}\label{eq:LLE_MI_NZ}
    \Tilde{P}^\pm = \frac{-(2+G)\phi_0 \pm \sqrt{\phi_0^2 -((2+G)^2-1)\alpha^2}}{2((2+G)^2-1)\gamma L}.
\end{equation}
Eqs.~(\ref{eq:LLE_MI_NZ}) and (\ref{eq:LLE_BS}) coincides for $G=0$, i.e. the ring cavity, where the instability of the homogeneous state coincides with the low-frequency limit of the modulationally unstable branch.
This peculiarity of FP resonator has been first pointed out in \cite{cole2018theory}.

\begin{figure}[t]
    \centering
     \includegraphics[width=8.6cm]{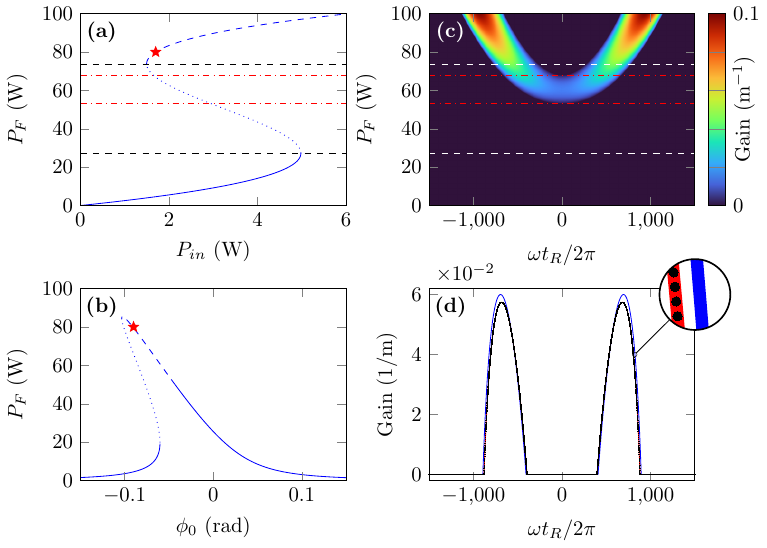}
     \caption{(a) Steady states as a function of pump power at $\phi_0=\SI{-0.09}{rad}$.
              (b) steady states as a function of linear phase at $P_{in}=\SI{1.7}{W}$.
              (c) Color level plot of gain $g(\omega)$ in the plane ($\omega$,$P_F$) of frequency and forward intracavity power, calculated from Eq.~(\ref{eq:FP_gain}).
              The dashed horizontal lines in (a) and (c) delimit the region of bistability, while dash-dotted horizontal lines delimit MI when $\omega_n \sim 0$.
              (d) Gain spectrum obtained from Eq.~(\ref{eq:FP_DR_general}) (black dots), from Eq.~(\ref{eq:FP_gain}) (red curve) and Eq.~(\ref{eq:FP_LLE_gain}) (blue curve) for an intracavity power $P_F=\SI{80}{W}$ and $\phi_0=\SI{-0.09}{rad}$; see red star in (a) and (b).
              Parameters: $\rho_1^2=\rho_2^2=0.98$ ($\mathcal{F}=156$), $\theta_1^2=\theta_2^2=0.02$, $\gamma = \SI{2}{W^{-1}/km}$, $\beta_1=c/1.5$, $\beta_2 = -\SI{20}{ps^2/km}$ and $L= \SI{0.1}{m}$.}
     \label{fig:SS_gS}
\end{figure}

In order to illustrate the results of linear stability analysis of models Eqs.~(\ref{eq:FP_2NLS}) and (\ref{eq:FP_LLE}), we consider for definiteness a fiber FP resonator, whose parameters are reported in Fig.~\ref{fig:SS_gS} caption.
Fig.~\ref{fig:SS_gS}(a,b) demonstrate examples of the intracavity steady state power obtained from Eq.~(\ref{eq:SS_power}) as a function of pump power and linear phase cavity, respectively.
The corresponding FP-LLE curves obtained from Eq.~(\ref{eq:LLE_SS}) are almost superimposed and they are not shown in order to make the figures more readable.
As usual, the nonlinear phase shift acquired by the intracavity field does not impact the resonance width, but it does tilt the resonance (see Fig.~\ref{fig:SS_gS}(b)).
If input power and cavity finesse are high enough, the resonances become increasingly tilted, resulting in a multivalued cavity response.
At certain values of $\phi_0$, the cavity can operate in a bistable regime, as shown in Fig.~\ref{fig:SS_gS}(a).
Fig.~\ref{fig:SS_gS}(c) displays the gain of MI calculated from Eq.~(\ref{eq:FP_gain}) for anomalous GVD regime, as a function of the mode frequencies and the intracavity forward field power $P_F$.
Also in this case the results obtained from FP-LLE Eq.~(\ref{eq:LLE_SS}) are practically identical (figure not shown).
Modulationally unstable steady states are represented by a dashed curve in Fig.~\ref{fig:SS_gS}(a) and Fig.~\ref{fig:SS_gS}(b).
Steady states which are unstable with respect to perturbations at zero frequency, corresponding to the negative-slope branch of the bistable response, are displayed in dotted curve.
One noteworthy characteristic of FP is that MI does not fully cover the CW unstable region, as it is the case for the ring cavity.
Fig.~\ref{fig:SS_gS}(d) exhibits an example of gain spectrum obtained from equations (\ref{eq:FP_DR_general}), (\ref{eq:FP_gain}) and (\ref{eq:FP_LLE_gain}).
The agreement between the different methods is perfect, even if the cavity's finesse is not very high.
This example shows that FP-LLE is a valuable tool for the description of MI in FP resonators.
However, the mean field model fails to describe some particular regimes, as it will be shown later.

\section{Pulsed pump}\label{sec:pulses}

\subsection{Stationary periodic solutions}
In this section we consider a pulsed pump with a repetition rate that matches the roundtrip time, meaning that $E_{in}(t+t_R)=E_{in}(t)$ is a periodic function. 
Equations~(\ref{eq:FP_2NLS}, \ref{eq:FP_BC}) can be analytically solved if dispersion is neglected ($\beta_2=0$)~\cite{firth1981stability}.
The solution can be written in implicit form as
\begin{align}\label{eq:solFBzt}
    \nonumber F(z,t)&=F(0,t-\beta_1 z)\exp\bigg[i\gamma|F(0,t-\beta_1 z)|^2z\\
    \nonumber &+i\gamma G\int_0^z|B(s,t-\beta_1 z+\beta_1s)|^2ds\bigg],\\
    \nonumber B(z,t)&=B(0,t+\beta_1 z)\exp\bigg[-i\gamma|B(0,t+\beta_1 z)|^2z\\
    &-i\gamma G\int_0^z|F(s,t+\beta_1 z-\beta_1s)|^2ds\bigg].
\end{align}
In the following we restrict our attention to a piecewise-constant (or quasi-CW) pump.
We thus consider a train of rectangular-shaped pump pulses of duration $\Delta t =f_r t_R<t_R$ and constant amplitude $E_{in0}$:
\begin{equation}\label{eq:squarepump}
    E_{in}(t)=\left
    \{\begin{array}{cc}
        E_{in0} & {\rm if }\;n\, t_R<t<n\, t_R+f_rt_R. \\
       0  & \rm{elsewhere }.
    \end{array}\right.
\end{equation}
Here, $f_r = \Delta t/t_R$ is the ratio between the pump pulse duration and the cavity roundtrip time (i.e. the duty cycle), and we search for time-periodic (steady-state) solutions.
In this case, Eqs.~(\ref{eq:solFBzt}) can be calculated explicitly~:
\begin{align}
    \nonumber F(z,t)=& F_0 \exp[i\gamma (|F_0|^2 z + G \cdot X_{F}(z,t))], \\
    \nonumber  &{\rm if }\; \beta_1 z+ n\,t_R<t<\beta_1 z+ n\, t_R+\Delta t; \\
    \nonumber  B(z,t) =& B_0 \exp[-i\gamma (|B_0|^2 z + G\cdot X_{B}(z,t))], \\
     & {\rm if }\; -\beta_1 z+ n\,t_R<t<-\beta_1 z+ n\, t_R+\Delta t, \label{eq:periodic_sol}
\end{align}
where $F_0$ and $B_0$ are complex constants to be determined, and 
\begin{align}
    \nonumber X_{F}(z,t) &= \int_0^z|B(s,t-\beta_1 z+\beta_1s)|^2ds,\\
    X_{B}(z,t) &= \int_0^z|F(s,t+\beta_1 z-\beta_1s)|^2ds.\label{eq:x_int}
\end{align}
The XPM terms $X_{F,B}$ are piece-wise linear functions in $(z,t)$.
Their expressions are rather cumbersome, and reported in Appendix~\ref{sec:pulses_exact} (Tables \ref{tab:table1} or \ref{tab:table2} depending whether  $f_r<0.5$ or $f_r>0.5$).
The complex constants of $F_0$ and $B_0$ are found by imposing boundary conditions Eqs.~(\ref{eq:FP_BC}).
We find that $F_0$ and $B_0$ are still given by Eqs.~(\ref{eq:FP_SS}), but with a different nonlinear phase shift~:
\begin{equation}\label{eq:phi_per}
    \phi_{NL} = \gamma (1 + f_r G)(1+\rho_2^2)|F_0|^2 L.
\end{equation}
We see that the effect of periodic pumping is to reduce the XPM by a factor $f_r$.
This is a peculiarity of the FP~: pumping the cavity with quasi-cw pulses does change the stationary states.
For a ring resonator this effect is absent because $G=0$.
If the pulse duration is much shorter than the roundtrip time ($f_r\ll 1$), the stationary states tends to the ones of ring cavity.

\subsection{Stability of quasi-CW solutions}
We now consider the stability of the periodic solutions Eqs.~(\ref{eq:periodic_sol}) with respect to dispersive perturbations ($\beta_2\neq0$).
We assume that the forward and the backward fields have the following form:
\begin{subequations}
    \begin{eqnarray}
        F(z,t) &=& F_p(z,t) (1 + f(z,t)) \;,\\
        B(z,t) &=& B_p(z,t) (1 + b(z,t)) \;,
    \end{eqnarray}
\end{subequations}
being $F_p, B_p$ the periodic solutions Eqs.~(\ref{eq:periodic_sol}), $f,b$ small perturbations and we insert this Ansatz in Eqs.~(\ref{eq:FP_2NLS}).
The inclusion of dispersion is not compatible with the discontinuous solutions Eqs.~(\ref{eq:periodic_sol}), so we approximate the square pulse with a smooth flat-top pulse with a rise-time which is much shorter than the pulse duration $\Delta t$, but long enough to neglect dispersive effect on the stationary periodic solution.
In practice, we neglect the terms $\beta_2\frac{\partial F_p}{\partial t}$ and $\beta_2\frac{\partial B_p}{\partial t}$ in the equations for the perturbations. Moreover, in the spirit of the approximation made in Sec.~\ref{sec:approx_sols}, we assume $G=0$ in the equations for the perturbations.
By expanding the perturbations as in Eqs.~(\ref{eq:perturbation_forms}), we obtain again Eqs.~(\ref{eq:MED}), with the steady state given by Eqs. (\ref{eq:FP_SS}, \ref{eq:phi_per}).
Eventually, the MI gain can be still calculated with Eq.~(\ref{eq:FP_gain}), which depends on the pulse duration through Eqs.~(\ref{eq:FP_SS}, \ref{eq:phi_per}).

\subsubsection{Mean field}
A similar analysis can be done also for the mean field model.
The steady periodic solution $\psi_p(t)$ of Eq.~(\ref{eq:FP_LLE}) with $\beta_2=0$ with square pulse pumping Eq.~(\ref{eq:squarepump}) has the same temporal shape of the pump, with peak power $P_p$ and constant phase.
The power $P_p$ is given by the following cubic equation~:
\begin{equation}
    \theta_1^2 P_{in} = P_p\left( \alpha^2 + (\phi_0 + 2\gamma L P_p(1 +f_rG))^2\right).
\end{equation}
Again, we see that the effect of the pulsed pumping is to reduce the XPM coefficient $G$ by a factor $f_r$.
For a CW pump $f_r=1$ and we recover Eq.~(\ref{eq:LLE_SS}).

We perform a linear stability by considering a perturbed solution of the form $\psi(t,\tau) = \psi_p(t) + \varepsilon_n(\tau)e^{i\omega_n t} + \varepsilon_{-n}(\tau)e^{-i\omega_n t}$ and include dispersion.
As done in the previous subsection, we approximate the square pulse with a smooth flat-top pulse with a rise-time much shorter than the pulse duration $\Delta t$, but long enough to minimise the dispersive effects on the stationary periodic solution.
This way, we can neglect the term $\beta_2\frac{\partial^2 \psi_p}{\partial t^2}$ in the equations for the perturbations.
By following the procedure described in Sec.~\ref{sec:FPLLE}, and assuming $\int_0^{\Delta t}e^{i\omega_n t}dt\approx 0$ ($n\neq0$), we find that the perturbations are ruled again by Eqs.~(\ref{eq:LLE_LS_LE}) with the substitution $G \rightarrow f_r G$. At the end, the results of the stability analysis given by Eqs.~(\ref{eq:FP_LLE_gain}, \ref{eq:LLE_MI_NZ}) are still valid with $G$ replaced by $f_r G$.

\begin{figure}[t]
    \centering
    \includegraphics[width=8.6cm]{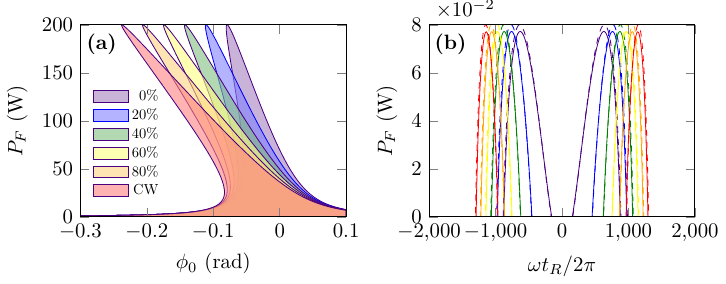}
    \caption{(a) Steady states as a function of $\phi_0$ for $P_{in}=4$ W and different values of $f_r$.
        (b) Gain spectrum obtained from Eq.~(\ref{eq:FP_gain}) (solid curves) and FP-LLE (dashed curves) for an intracavity power $P_F=90$~W and $\phi_0=-0.04$~rad for different values of $f_r$. Cavity parameters: see Fig.~\ref{fig:SS_gS}.} 
    \label{fig:DutyCycle_Effect}
\end{figure}

To illustrate the effect of the duration of the pump pulses, we consider the FP cavity used in Fig.~\ref{fig:SS_gS} with pulsed pumping. Fig.~\ref{fig:DutyCycle_Effect}(a) presents the cavity response plotted as function of linear phase for various pulse durations.
It can be observed that as the pulse duration increases, the resonance shape becomes more tilted.
This phenomenon is not observed in ring cavities since the XPM effect is absent, which results in an unchanged resonance shape.
This observation indicates that the pulse duration is a significant control parameter in FP cavities. To further emphasize this relationship, Fig.~\ref{fig:DutyCycle_Effect}(b) shows the MI gain spectrum for different pulse durations while maintaining the intracavity power constant. Interestingly, it is noticed that the maximum gain remains constant regardless of pulse duration, but the corresponding frequency is dependent on it.

\begin{figure}[t]
    \includegraphics[width=8.6cm]{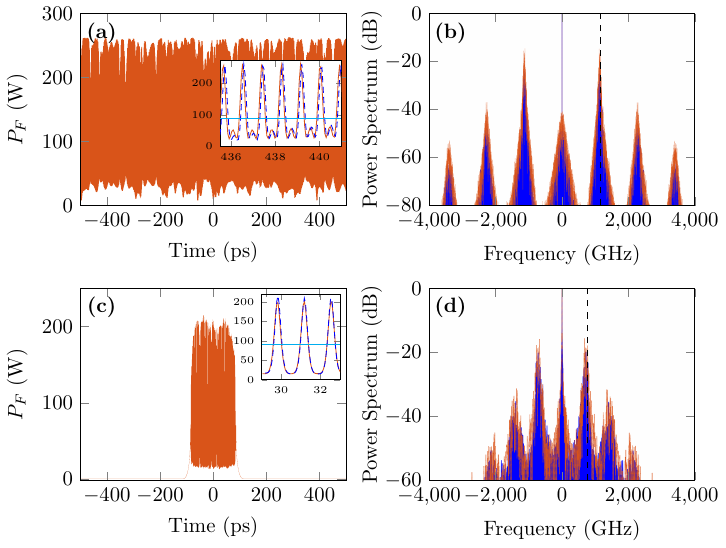}
    \caption{
        Numerical simulations of MI-induced frequency comb generation for CW (a,b) and pulsed pumping with $f_r=0.2$ (c,d).
        Intracavity power (a,c) and spectrum (b,d) after $1000$ roundtrips.
        Orange curves correspond to FP-LLE, while blue curves correspond to the full model.
        Horizontal cyan line in the insets represents the input field.
        Black vertical dashed lines indicate the peak MI gain from Eq.~(\ref{eq:FP_LLE_gmax}): $\SI{1148}{GHz}$ in (b) and $\SI{759}{GHz}$ in (d).
        Initial intracavity power $P_F=\SI{90}{W}$ and $\phi_0=\SI{-0.04}{rad}$, rest of parameters as in Fig.~\ref{fig:SS_gS}.}
    \label{fig:prop_DutyCycle_Effect}
\end{figure}

The results of linear stability analysis permit to predict the position of the unstable spectral bands even in the fully nonlinear regime, where an almost periodic train of pulses, i.e. a frequency comb, is generated.
Figure~\ref{fig:prop_DutyCycle_Effect} show the results of numerical solution of FP-LLE with a standard Fourier split-step method.
The initial condition is a CW (a,b) or a periodic steady-state (c,d) perturbed by a small random noise and it is propagated over $1000$ roundtrips in order to reach a stable state.
For a CW pump, we see in Fig.~\ref{fig:prop_DutyCycle_Effect}(a) that the field fills all the cavity (the time window extends from $-t_R/2$ to $t_R/2$) and is composed of a quasi periodic sequence of short pulses (see inset).
The spectrum is composed of several lines generated by cascaded FWM, and the position of the first sideband is perfectly predicted by the LSA (black dashed line).
For a pulsed pump, we see in Fig.~\ref{fig:prop_DutyCycle_Effect}(c) that the cavity is partially empty.
The field is composed of bursts of short pulses, as highlighted in the inset.
The spectrum is still composed of several lines, but the spacing is different as predicted by the LSA (black dashed line).
In Fig.~\ref{fig:prop_DutyCycle_Effect}(a-d) orange curves are the temporal and spectral traces obtained from the numerical solution of FP-LLE, while blue curves are obtained from coupled NLSE.
For the numerical solution of coupled NLSE Eqs.~(\ref{eq:FP_2NLS}, \ref{eq:FP_BC}) we used a \textit{split-step, predictor-corrector} method evolved in time~\cite{sun2019stable}.
We can see a very good agreement of the spectra in panels (b,d).
The slight discrepancies are mainly due to the fact that the initial seed is random noise, which is not identical in the two simulations.
The overall agreement of temporal traces in panels (a,c) is also good.
The insets shows a zoom on a limited temporal span, showing a remarkable quantitative agreement.
The numerical simulation of FP-LLE took only $0.5$ minutes on a standard workstation, while the full model took $9$ hours ($1000$ times slower) for the same number of roundtrips and the same frequency span.
The long computation time for the coupled NLSE is mainly caused by the counter-propagation, which imposes to solve two equation with two different group velocities.
The experimental demonstration of these phenomena will be published elsewhere~\cite{bunel2023impact}.

\section{Limitations of the mean field model}\label{sec:results}

The examples presented in the previous sections showed that FP-LLE permits to accurately reproduce the results of the full model.
However, the derivation of FP-LLE involves approximations that result in inherent limitations.
In order to identify the regions in the parameters space where the mean-field model breaks down, we draw a chart of instabitity from the results of LSA.

\begin{figure}[t]
    \centering 
    \includegraphics[width=8.6cm]{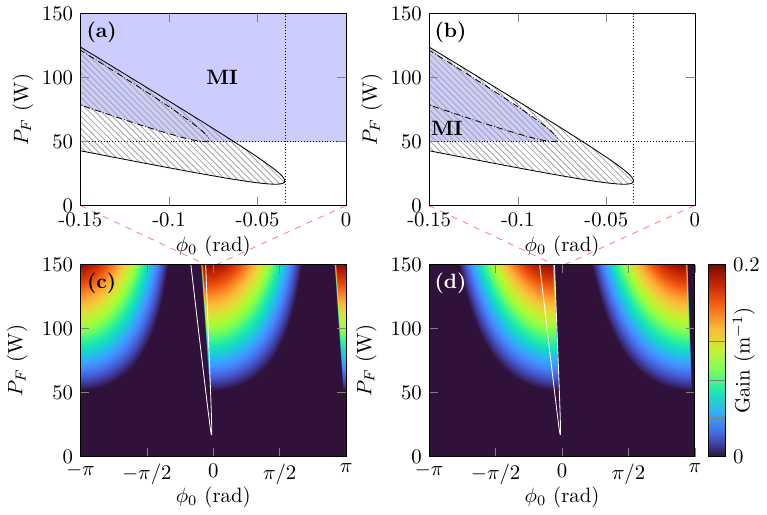}
    \caption{
        Instability chart, obtained using FP-LLE model (a,b), and the full model (c,d) in ($\phi_0$, $P_F$) plane for (a,c) anomalous and (b,d) normal GVD regimes.
        Modulationally unstable domains are shaded in blue and continuous wave unstable domains are hatched.
        Solid curves correspond to $P^\pm$, which delimit the bistable region.
        Dash-dotted curves delimit the low-frequency limit of the MI unstable domains from Eqs.~(\ref{eq:LLE_MI_NZ}).
        Vertical dotted lines separate mono and bistable regimes and the horizontal dotted lines show the threshold power $P_{th}$.
        Panels (c) and (d) show the two-dimensional map of the maximum gain for the full model from Eq.~(\ref{eq:FP_gain}).
        Parameters as in Fig.~\ref{fig:SS_gS}.}
    \label{fig:TuringDomain}
\end{figure}

We start by considering FP-LLE.
Modulation instability occurs when $g_{max}>0$, and using Eq.~(\ref{eq:FP_LLE_gmax}), we can derive the intracavity power threshold $P_{th} = \alpha/(2\gamma L)$, which is independent of $\phi_0$ and the sign of the GVD parameter.
Figure~\ref{fig:TuringDomain} illustrates the bistability and MI regions as a function of the cavity linear phase $\phi_0$, with hatched areas corresponding to the negative slope branch of the bistable curve between $P^-$ and $P^+$ from Eq.~(\ref{eq:LLE_BS}), blue area to the MI region, and white areas to the stable region.
For $\beta_2<0$, MI arises in both bistable and monostable regimes when $P_F >P_{th}$, whereas for $\beta_2>0$, MI arises only in the bistable regime and is confined to a relatively small domain.
The dash-dotted curve delimits the low-frequency limit of the MI unstable domains from Eqs.~(\ref{eq:LLE_MI_NZ}).
It is worth noting that for certain values of ($\phi_0, P_F)$ in the bistable region, only the mode $\omega=0$ is unstable, which is a distinguishing feature of FP cavities.

\begin{figure}[b]
    \centering
    \includegraphics[width=8.6cm]{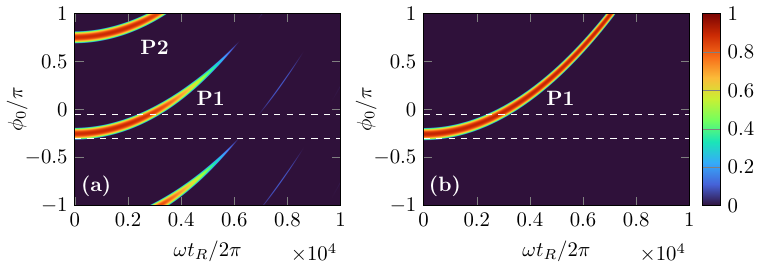}
    \caption{
        Color level plot of gain $g(\omega)$ in the plane ($\omega, \phi_0$) of frequency and cavity linear phase (detuning), calculated from (a) Eq.~(\ref{eq:FP_gain}) and (b) FP-LLE Eq.~(\ref{eq:FP_LLE_gain}), with intracavity power $P_F=500$ W.
        The dashed horizontal lines delimit an estimate of the region of validity of the LLE.  Parameters as in Fig.~\ref{fig:SS_gS}.}    
    \label{fig:pcolor_comparison}
\end{figure}

For the full model, the boundary between the stable and unstable regions are given by Eq.~(\ref{eq:MI_thresholds}), which is the solution of the equation $g(\omega)=0$ from Eq.~(\ref{eq:FP_gain}).
Figures~\ref{fig:TuringDomain}(c,d) show the MI gain calculated from (\ref{eq:FP_gain}), over the full range of cavity linear phase $(-\pi,\pi)$.
Differently than FP-LLE, the MI power threshold does depend on $\phi_0$.
In particular, both in the normal and the anomalous regimes, we can see two unstable tongues, one centered around zero detuning and the other around $\phi_0=\pi$.
The unstable region centered at $\phi=0$ corresponds to even values of $m$ in Eq.~(\ref{eq:str}), meaning that the unstable frequencies corresponds to cavity resonance $\omega t_R=2n\pi$.
Whereas, the unstable region centered at $\phi=\pi$ corresponds to odd values of $m$ in Eq.~(\ref{eq:str}), meaning that the unstable frequencies $\omega t_R=(2n+1)\pi$ are in between two resonances (anti-resonance).
We may identify in this second case the period-doubling (P2) MI, which has been described before for ring cavities \cite{conforti2016parametric,haelterman1992period,coen1997modulational,bessin2019real}.
The difference between standard (i.e. period one, P1) MI and P2-MI is that the modulations developing from the instability are in phase (P1) or shifted by half a temporal period (P2) at each roundtrip.
It is worth noting that previous theoretical studies on P2-MI were based on the Ikeda map, which does not permit to resolve the cavity modes.
As happens in ring resonators, the FP-LLE fails to predict P2 instabilities.
To highlight this feature, we report in Fig.~\ref{fig:pcolor_comparison}a, the gain in ($\omega$, $\phi_0$) obtained from (\ref{eq:FP_gain}). The two instability branches, are labelled P1 and P2 on the figure.
The P1 instability is captured by FP-LLE, as shown in Fig.~\ref{fig:pcolor_comparison}b.
On the other hand, the P2 instability is not visible in the gain calculated from FP-LLE. 

\begin{figure}[t]
    \includegraphics[width=8.6cm]{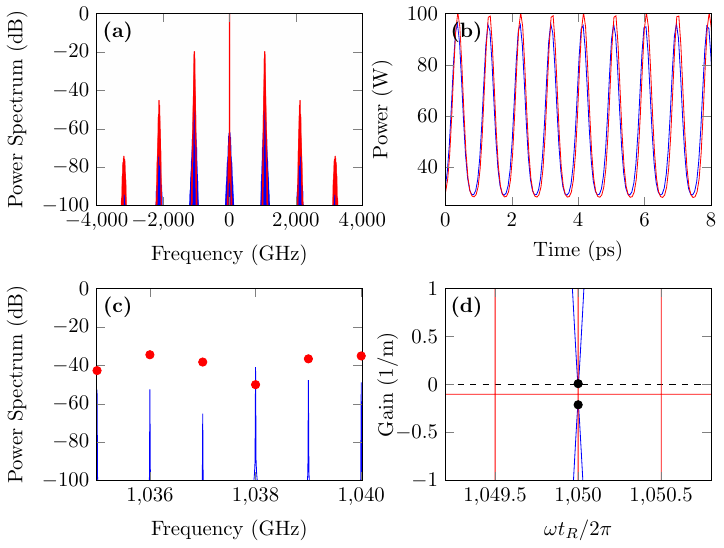}
    \caption{
        (a) Output spectrum after $10000$ roundtrips.
        Vertical dashed line is the maximally unstable frequency obtained from LSA.
        Blue and red curves correspond to the full model and FP-LLE (b) Intracavity power at the output mirror as a function of normalised time $t-nt_R$, $n=10000$.
        Blue and red curves correspond to the full model and FP-LLE.
        (c) Zoom on the spectrum around the $1038$-th resonance (FSR=$\SI{1}{GHz}$).
        Blue curve and red dots correspond to the full model and FP-LLE (d)  Graphical solution of Eq.~(\ref{eq:FP_DR_general}).
        $P_F=\SI{50}{W}$, $\phi_0=0$.
        Rest of parameters as in Fig.~\ref{fig:SS_gS}.
        Simulation time around $23$ hours for the full model, $50$ seconds for FP-LLE.}
    \label{fig:P1}
\end{figure}

In the following we complement the results of the linear stability analysis with numerical solution of the governing equations in the fully developed nonlinear regime.
Figure~\ref{fig:P1} reports the generation of a P1-MI comb from numerical simulations of Eqs.~(\ref{eq:FP_2NLS},\ref{eq:FP_BC}) (blue curves) and FP-LLE Eq.~(\ref{eq:FP_LLE}) (red curves).
Figure \ref{fig:P1}(a) shows the output spectrum after $10000$ roundtrips, where a steady state is reached.
The position of the unstable bands is well predicted by LSA ($f_{max}=\SI{1050}{GHz}$).
Figure~\ref{fig:P1}(b) shows a zoom of the the temporal behaviour of intracavity field at the output mirror at roundtrip $10000$.
The field is composed of an almost periodic train of short pulses, which reproduces itself at each roundtrip.
Figures~\ref{fig:P1}(a,b) show a good agreement between the full and the mean-field model (blue and red curves).
For this simulation, the computation time for the mean field model was divided by around $1500$ times with respect to the full model.
Figure~\ref{fig:P1}(c) shows a zoom of the spectrum around the maximum of the first band: only frequencies corresponding to the the cavity resonances are excited, as predicted by LSA showed in Figure~\ref{fig:P1}(d).

\begin{figure}[t]
    \includegraphics[width=8.6cm]{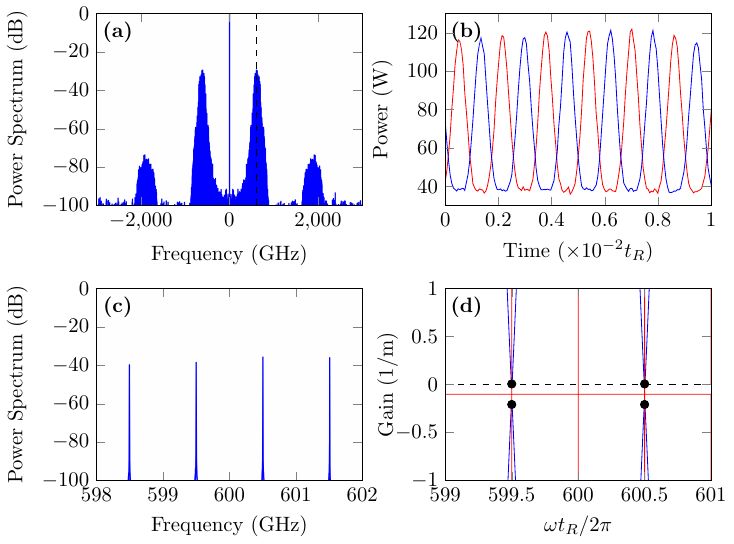}
    \caption{
        (a) Output spectrum after $10000$ roundtrips.
        Vertical dashed line is the maximally unstable frequency obtained from LSA.
        (b) Intracavity power at the output mirror as a function of normalised time $(t-nt_R)/t_R$, $n=9999,10000$.
        Blue and red curves correspond to the roundtrip $9999$ and $10000$.
        (c) Zoom on the spectrum around the $600$-th resonance (FSR=$\SI{1}{GHz}$).
        (d) Graphical solution of Eq.~(\ref{eq:FP_DR_general}).
        $P_F=\SI{50}{W}$, $\phi_0=0.98\pi$.
        Rest of parameters as in Fig.~\ref{fig:SS_gS}.
        Simulation time around $20$ hours.}
    \label{fig:P2}
\end{figure}

Figure~\ref{fig:P2} reports the generation of a P2-MI comb from numerical simulations of Eqs.~(\ref{eq:FP_2NLS}, \ref{eq:FP_BC}).
Figure~\ref{fig:P2}(a) shows the output spectrum after $10000$ roundtrips, where a steady state is reached.
The position of the unstable bands is well predicted by LSA ($f_{max}=\SI{555}{GHz}$).
Quite surprisingly, the first FWM band around $2f_{max}$ is not generated, whereas is clearly visible the second FWM band around $3f_{max}$.
Figure~\ref{fig:P2}(c) shows a zoom of the spectrum around the maximum of the first band~: the modes have frequencies which fall in between two adjacent cavity resonances, as predicted by LSA showed in Figure~\ref{fig:P2}(d).

This observation may explain why first order FWM is not present.
Indeed, the spectrum of the field is composed of lines at anti-resonance.
Frequency doubling of the first sideband will lead to lines at cavity resonances, which are inhibited in this configuration.
The analysis of the features of the fully developed P2-MI pattern is still under investigation.
Figure~\ref{fig:P2}(b) shows the the temporal behaviour of intracavity field at the output mirror at two consecutive roundtrips.
We clearly see that the two traces are out-of-phase, a the typical signature of P2-MI.

\section{Conclusion}\label{sec:conclusion}

We have studied modulation instability in Kerr Fabry-Perot cavities.
Starting from a coupled NLSE description of the cavity dynamics, we have derived the exact dispersion relation for the perturbations and we found approximate analytical expressions for the instabilities threshold and gain spectrum of modulation instability.
We showed that, in contrast to ring-resonators, both the stationary solutions and the gain spectrum depends on the pump-pulse duration.
We derived the extended Lugiato-Lefever equation for the Fabry-Perot resonator (FP-LLE) starting from coupled nonlinear Schr\"odinger equations (rather than Maxwell-Bloch equations as done in~\cite{cole2018theory}) and we compared the results of the stability analysis of the two models.
While FP-LLE gives overall good results, we showed regimes that are not captured by the mean-field limit, namely the period-two modulation instability, which may appear in highly detuned or nonlinear regimes.
We reported numerical simulations of the generation of MI-induced Kerr combs by solving FP-LLE and the coupled NLSE.
Overall, our study aims at gaining a deeper understanding of the nonlinear dynamics of Fabry-Pérot cavities, which could have important implications for the development of new technologies and applications in fields such as telecommunications, optical sensing, and metrology.
The findings of our study could potentially assist the design of more efficient and robust cavity-based systems.

\begin{acknowledgments}
    The present research was supported by the Agence Nationale de la Recherche (Programme Investissements d’Avenir, I-SITE VERIFICO) and IRCICA.
\end{acknowledgments}

\appendix
\section{Modulation instability in a ring cavity with a tap coupler} \label{sec:ringMI}

We consider a ring resonator composed of two spans of identical fiber connected to an input coupler $1$ and a tap coupler $2$, as illustrated in Fig. \ref{fig:ring_resonator}.
If there is no coupling between forward and backward fields, it is easy to obtain a map wich describes the behavior of the system at each roundtrip \cite{zezyulin2011modulational}.
The fields propagating in the two spans satisfies NLSE~:
\begin{align}
    i\frac{\partial F_n}{\partial z} \label{eq:propFring}
    -\frac{\beta_2}{2}\frac{\partial^2 F_n}{\partial t^2}+\gamma|F_n|^2F_n &=0,\; 0<z<L,\\
    i\frac{\partial B_n}{\partial z} \label{eq:propBring}
    -\frac{\beta_2}{2}\frac{\partial^2 B_n}{\partial t^2}+\gamma|B_n|^2B_n &=0,\; L<z<2L,
\end{align}
and they are coupled by the following boundary conditions at couplers:
\begin{align}
    F_{n+1}(0,t)&=\theta_1 E_{in} + \rho_1 e^{i\phi_0} B_n(2L,t),\label{eq:BC1ring}\\
    B_n(L,t)&= \rho_2 F_n(L,t).\label{eq:BC2ring}
\end{align}
The total linear phase $\phi_0$ accounts for propagation and phase from the couplers and the index $n$ counts the  number of roundtrips.

\subsection{Steady states}
Steady state solutions of Eqs.~(\ref{eq:propFring}-\ref{eq:propBring}) reads as
\begin{align}
    F_n(z,t)&=F_0e^{i\gamma P_F z},\;P_F=|F_0|^2,\\
    B_n(z,t)&=B_0e^{i\gamma P_B z},\;P_B=|B_0|^2.
\end{align}
By using the boundary conditions, we find the cavity transfer function:
\begin{equation}\label{eq:steady_ring}
    F_0=\frac{\theta_1 E_{in}}{1-\rho_1\rho_2\exp[i(\phi_0+\phi_{NL})]},
\end{equation}
which permits to write the input power $P_{in}=|E_{in}|^2$ as a function of intracavity forward power $P_F=|F_0|^2$ as:
\begin{equation}
    P_{in}=\frac{P_F}{\theta_1^2}\left(1+(\rho_1\rho_2)^2-2\rho_1\rho_2\cos(\theta_0)\right),
\end{equation}
with $\theta_0=\phi_0+\phi_{NL}=\phi_0+\gamma P_F L(1+\rho^2_2)$.

It is worth noting that Eq.~(\ref{eq:steady_ring}) is equivalent to the steady-state of a FP resonator with $G=0$ and it is also equivalent to the steady state of a ring resonator of length $2L$ if $\rho_2=1$.

\subsection{Linear stability analysis}
We consider a perturbation of the steady state in the following form
\begin{align}
    F_n(z,t)&=(\sqrt{P_F}+\eta)e^{i\gamma P_F z},\\
    B_n(z,t)&=(\rho_2\sqrt{P_F}+\varepsilon)e^{i\gamma \rho_2^2 P_F z}e^{i\gamma \phi_B},
\end{align}
where we have assumed without loss of generality $F_0$ real, which fixes the phase $\phi_B=\gamma LP_F(1-\rho_2^2)$ through boundary condition (\ref{eq:BC2ring}).
Linearization around steady solutions gives the equations for the perturbations:
\begin{align}
    \label{eq:pert_eta_map} i\eta_z-\frac{\beta_2}{2}\eta_{tt}+\gamma P_F(\eta+\eta^*)&=0,\\ 
    \label{eq:pert_eps_map} i\varepsilon_z-\frac{\beta_2}{2}\varepsilon_{tt}+\gamma \rho_2^2P_F(\varepsilon+\varepsilon^*)&=0. 
\end{align}
 
We split perturbations into real and imaginary parts, $\eta=a+ib$ and $\varepsilon=c+id$ , we substitute into Eqs.~(\ref{eq:pert_eta_map}-\ref{eq:pert_eps_map}) and Fourier transform to get
\begin{align}\label{eq:evopert}
    \nonumber\begin{pmatrix}
    \hat a \\ \hat b
    \end{pmatrix}_z &=
    \begin{pmatrix}
    0 &-\frac{\beta_2\omega^2}{2}\\
    \frac{\beta_2\omega^2}{2}+2\gamma P_F & 0
    \end{pmatrix}
    \begin{pmatrix}
    \hat a \\ \hat b
    \end{pmatrix},\\
    \begin{pmatrix}
    \hat c \\ \hat d
    \end{pmatrix}_z &=
    \begin{pmatrix}
    0 &-\frac{\beta_2\omega^2}{2}\\
    \frac{\beta_2\omega^2}{2}+2\gamma \rho_2^2P_F & 0
    \end{pmatrix}
    \begin{pmatrix}
    \hat c \\ \hat d
    \end{pmatrix}.
\end{align}
The fundamental matrix solutions of systems Eqs.~(\ref{eq:evopert}) are
\begin{align}
    M(z) &=
    \begin{pmatrix}
    \cos kz &-\frac{\beta_2\omega^2}{2k}\sin kz\\
    \frac{2k}{\beta_2\omega^2}\sin kz & \cos kz
    \end{pmatrix},\\
    N(z) &=
    \begin{pmatrix}
    \cos k_\rho z &-\frac{\beta_2\omega^2}{2k\rho}\sin k_\rho z\\
    \frac{2k_\rho}{\beta_2\omega^2}\sin k_\rho z & \cos k_\rho z
    \end{pmatrix},\;\;
\end{align}
with $k,k_\rho$ defined in Eqs.~(\ref{eq:k_omega}) with $P=P_F$ and $\lambda=i\omega$.
The boundary conditions give the following relations:
\begin{align}\label{eq:BC_abcd}
    \nonumber\begin{pmatrix}
    \hat c_n(L) \\ \hat d_n(L)
    \end{pmatrix}&=
    \rho_2
    \begin{pmatrix}
    \hat a_n(L) \\ \hat b_n(L)
    \end{pmatrix},\\
    \begin{pmatrix}
    \hat a_{n+1}(0) \\ \hat b_{n+1}(0)
    \end{pmatrix}&=
    \rho_1
    \begin{pmatrix}
    \cos \theta_0 & -\sin\theta_0\\
    \sin\theta_0 & \cos\theta_0
    \end{pmatrix}
    \begin{pmatrix}
    \hat c_n(2L) \\ \hat d_n(2L)
    \end{pmatrix}.
\end{align}

By combining propagation and boundary conditions, we get the following difference equation:
\begin{equation}
    \begin{pmatrix}
    \hat a_{n+1}(0) \\ \hat b_{n+1}(0)
    \end{pmatrix}=S
    \begin{pmatrix}
    \hat a_{n}(0) \\ \hat b_{n}(0)
    \end{pmatrix},\;\; S=\rho_1\rho_2 R N(L) M(L),
\end{equation}
and $R$ is the rotation matrix defined in Eq.~(\ref{eq:BC_abcd})

The eigenvalues $\lambda_{1,2}$ of matrix $S$ determines the stability of the steady solution.
We find
\begin{equation}
    \lambda_{1,2}=\frac{\tilde\Delta}{2}\pm\sqrt{\frac{\tilde\Delta^2}{4} - |\rho_1\rho_2|^2},
\end{equation}
with $\tilde\Delta$ as defined in Eq.~(\ref{eq:FP_gain}).
Instability takes places if $|\lambda_{1,2}|>1$ and the MI gain is
\begin{equation}\label{eq:gain_map}
    g(\omega)=\frac{1}{2L}\ln\max\left|\frac{\tilde\Delta}{2}\pm\sqrt{\frac{\tilde\Delta^2}{4} - |\rho_1\rho_2|^2}\right|,
\end{equation}
which coincides with the gain for the FP resonator found before in Eq.~(\ref{eq:FP_gain}).

\section{FP-LLE derivation} \label{sec:LLE}
We derive a mean field model, which generalises the Lugiato-Lefever equation, for the description of a passive driven fiber Fabry-Perot cavity.
We follow an approach similar to the one developed in Ref.~\cite{cole2018theory} but with a different starting point, namely coupled NLS  [Eqs.~(\ref{eq:FP_2NLS}, \ref{eq:FP_BC})] rather than Maxwell-Bloch equations.
The main steps are~: (i) change variables to make the boundary conditions periodic and to include the pump term in the propagation equation; (ii) take the good-cavity (or mean field) approximation; (iii) derive a partial differential equation using the modal equations.
We start by defining the following change of variables~\cite{lugiato1988nonlinear,lugiato2015nonlinear}~:
\begin{subequations}\label{eq:transf}
    \begin{eqnarray}
        \nonumber \tilde{F}(z,t) &=& \exp\left[\frac{z-L}{L}\left(\ln{\rho_1}+i\frac{\phi_0}{2}\right)-\sigma z\right]F(z,t)\\
       &+& \frac{\theta_1}{\rho_1}\exp\left(-i\frac{\phi_0}{2}\right) \frac{z-L}{2L}E_{in}(t-\beta_1 z)\;,\\
        \nonumber \tilde{B}(z,t) &=& \exp\left[-\frac{z}{L}\left(\ln{\rho_2}+i\frac{\phi_0}{2}\right)-\sigma z-i\frac{\phi_0}{2}\right ]B(z,t)\\
         &-& \frac{\theta_1}{\rho_1}\exp\left(-i\frac{\phi_0}{2}\right) \frac{z-L}{2L}E_{in}(t+\beta_1 z),
\end{eqnarray}
\end{subequations}
with $\sigma = \frac{1}{2L} \ln(\rho_1/\rho_2)$.
This transformation is more general than the one proposed in~\cite{lugiato1988nonlinear,lugiato2015nonlinear} because we allow the two mirrors to be different and the pump may vary in time.
The boundary conditions given by Eqs.~(\ref{eq:FP_BC}) for the new variables are simplified to~:
\begin{equation}\label{eq:FP_BC_nv}
\begin{split}
    \tilde{F}(0,t) =& \tilde{B}(0,t),\\
    \tilde{F}(L,t) =& \tilde{B}(L,t)
\end{split}
\end{equation}
The simplification of the boundary conditions is payed by an increase in complexity of the propagation equations.
We thus restrict our analysis to good cavities ($\rho_{1,2}\rightarrow 1$ and $\phi_0\rightarrow 0$), for which we can obtain a mean field description.
From Eqs.~(\ref{eq:transf}) we calculate $\partial_z F, \partial_t F, \partial_z B ,\partial_t B$ as a function of $\tilde F,\tilde B$ and their derivatives.
We truncate the obtained expressions at first order in $\rho_{1,2}$ and $\phi_0$ and insert them into Eqs.~(\ref{eq:FP_2NLS}).
By considering that dispersion and nonlinearity are weak (assumptions already used to derive NLS ), we can use zero order expansion ($\tilde F=F$, $\tilde B=B$) in the dispersive and nonlinear terms.
These approximations permit to greatly simplify the propagation equations as follows~:
\begin{subequations}\label{eq:NLS_nv}
    \begin{align}
        \nonumber \frac{\partial \tilde{F}}{\partial z} + \beta_1 \frac{\partial \tilde{F}}{\partial t} + i\frac{\beta_2}{2}\frac{\partial^2 \tilde{F}}{\partial t^2}-\frac{1}{L}\left( \ln{\rho_1\rho_2}+i\frac{\phi_0}{2} \right)\tilde{F} \\
        -\frac{\theta_1}{2L} E_{in}(t-\beta_1 z) = i\gamma \left( |\tilde{F}|^2 + G|\tilde{B}|^2 \right)\tilde{F}, \\
        \nonumber  -\frac{\partial \tilde{B}}{\partial z} + \beta_1 \frac{\partial \tilde{B}}{\partial t} + i\frac{\beta_2}{2}\frac{\partial^2 \tilde{B}}{\partial t^2}-\frac{1}{L}\left( \ln{\rho_1\rho_2}+i\frac{\phi_0}{2} \right)\tilde{B} \\
        -\frac{\theta_1}{2L} E_{in}(t+\beta_1 z) = i\gamma \left( |\tilde{B}|^2 + G|\tilde{F}|^2 \right)\tilde{B}.
    \end{align}
\end{subequations}

\subsection{Modal equations}
We start by finding the modes of the empty and undriven (cold) cavity, then we expand the fields of the hot cavity in terms of the modes of the cold cavity and derive the equations ruling the slow evolution of the modal amplitudes. 
By taking $\beta_2=\phi_0=E_{in}=\gamma=0$, we solve Eqs.~(\ref{eq:NLS_nv}) with boundary conditions Eqs.~(\ref{eq:FP_BC_nv}), to find
\begin{subequations}
    \begin{eqnarray}
        \tilde{F}(z,t) &=&  A \exp\left[ \left(\beta_1\lambda + \frac{\ln{\rho_1\rho_2}}{2L} \right)z\right] e^{-\lambda t}, \\
        \tilde{B}(z,t) &=&  A \exp\left[-\left(\beta_1\lambda + \frac{\ln{\rho_1\rho_2}}{2L}\right)z \right] e^{-\lambda t},
    \end{eqnarray}
\end{subequations}
with
\begin{equation}\label{eq:boundary}
    \exp[2\beta_1\lambda L + \ln(\rho_1\rho_2)] = 1,
\end{equation}
where $A$ and $\lambda$ are constants.
By defining $\lambda=\kappa+i\omega$, we get from Eq.~(\ref{eq:boundary})
\begin{equation}
    \omega_m = \frac{m\pi}{\beta_1 L}\;,\quad \text{and} \quad \kappa = -\frac{\ln(\rho_1\rho_2)}{2\beta_1 L},
\end{equation}
which are the frequencies and the decay rate of the cavity modes.
We may write the modes of the cold cavity as 
\begin{subequations}
    \begin{eqnarray}
        \tilde{F}_m(z,t) &=&  e^{-\kappa t}e^{-i\omega_m(t-\beta_1 z)}\\
        \tilde{B}_m(z,t) &=&  e^{-\kappa t}e^{-i\omega_m(t+\beta_1 z)}.
    \end{eqnarray}
\end{subequations}
The fields in the full model can now be written as the sum of the loss-less cold cavity modes, allowing for a slow temporal variation of the modal amplitudes, which is induced by pumping, nonlinear and dispersive effects.
Note that the small damping  $\kappa$ is also accounted for in the slowly varying modal amplitudes.
We thus may write:
\begin{subequations}\label{eq:apdx_field}
    \begin{eqnarray}
        \tilde{F}(z,t) &=&  \sum_m a_m(t) e^{-i\omega_m(t-\beta_1 z)}\\
        \tilde{B}(z,t) &=&  \sum_m a_m(t) e^{-i\omega_m(t+\beta_1 z)}.
    \end{eqnarray}
\end{subequations}
We consider a periodic input, synchronised with the cavity repetition rate,  which can be expanded in Fourier series as follows
\begin{equation}\label{eq:apdx_pump}
    E_{in}(t) = \sum_m S_m e^{-i\omega_m t}.
\end{equation}
We insert Eq.~(\ref{eq:apdx_field}) and Eq.~(\ref{eq:apdx_pump}) in (\ref{eq:NLS_nv})a, multiply by $e^{i\omega_n (t-\beta_1 z)}$ and integrate in $z \in [-L,L]$, to obtain :
\begin{align}\label{eq:modal}
    \nonumber &\beta_1 \dot{a_n} - \left(\frac{\ln(\rho_1\rho_2)}{2L}+i\frac{\phi_0}{2L}\right)a_n \\
    \nonumber &+ i\frac{\beta_2}{2} \left(\ddot{a}_n -2i \omega_n \dot{a}_n -\omega_n^2 a_n\right) - \frac{\theta_1}{2L} S_n = \\
    &i\gamma \sum_{n',n''} a_{n'}a_{n''}^* (a_{n-n'+n''} + G a_{n+n'-n''}e^{-2i(\omega_{n'}-\omega_{n''})t}).
\end{align}
We assume that the modal amplitudes change slowly over a roundtrip, i.e. $|\dot{a}_n| \ll |\omega_n a_n|$.
This assumption permits to simplify the dispersive contribution, by neglecting the time derivatives of the modal amplitudes in the third term of Eq.~(\ref{eq:modal}). Moreover, by integrating Eq.~(\ref{eq:modal}) in time over one roundtrip, and considering $a_n(t)$ constant in this range,  the fast oscillations in the second nonlinear term are averaged out. We eventually obtain~:
\begin{align}
    \nonumber \dot{a}_n + \left(\kappa -i\frac{\phi_0}{2\beta_1 L} -i\frac{\beta_2}{2\beta_1}\omega_n^2\right)a_n 
    - \frac{\theta_1}{2\beta_1 L} S_n = \\
    i\frac{\gamma}{\beta_1} \left( \sum_{n',n''}a_{n'}a_{n''}^*a_{n-n'+n''} + G a_n \sum_{n'} |a'_n|^2 \right).
\end{align}
The same equation is also obtained by following a similar procedure starting from (\ref{eq:NLS_nv})b.

\subsection{Mean field FP-LLE}
We may now define the slowly varying envelope of the forward and backward fields in the laboratory frame as
\begin{subequations}
    \begin{eqnarray}
        \psi(z,t) &=& \sum_m a_m(t) e^{-i\omega_m t}e^{i\beta_1 \omega_m z} \\
        \psi_B(z,t) &=& \sum_m a_m(t) e^{-i\omega_m t}e^{-i\beta_1 \omega_m z}.
    \end{eqnarray}
\end{subequations}
It is apparent that the fields are periodic in space of period $2L$ and they satisfy $\psi(z, t) = \psi_B(-z, t)$.
Thanks to this relation we can relate the fields in the ’nonphysical’ cavity $-L < z < 0$ to the real cavity $0 < z < L$ to their conter-propagating counterparts~\cite{cole2018theory}.
By using
\begin{eqnarray*}
    \frac{\partial \psi}{\partial t} &=& \sum_m (\Dot{a}_m(t) -i\omega_m a_m) e^{-i\omega_m t}e^{i\beta_1 \omega_m z} \;, \\
    \frac{\partial^n \psi}{\partial z^n} &=& \sum_m (i\beta_1 \omega_m)^n a_m(t) e^{-i\omega_m t}e^{i\beta_1 \omega_m z}\;,
\end{eqnarray*}
we easily get
\begin{align}\label{eq:FP_LLE_t1}
    \nonumber \frac{\partial \psi}{\partial t} &+ \frac{1}{\beta_1} \frac{\partial \psi}{\partial z} + \left(\kappa -i\frac{\phi_0}{2\beta_1 L}\right)\psi\\
    \nonumber &+i\frac{\beta_2}{2\beta_1^3}\frac{\partial^2\psi}{\partial z^2} - \frac{\theta_1}{2\beta_1 L}E_{in}(t-\beta_1 z) \\
    &= i\frac{\gamma}{\beta_1} \left( |\psi|^2 + \frac{G}{2L} \int_{-L}^L |\psi(z',t)|^2 dz'\right)\psi.
\end{align}
By means of the change of variable $z \rightarrow -z + t/\beta_1$ [mod $2L$] and multiplying by the roundtrip time $t_R = 2\beta_1L$ we get~:
\begin{align}\label{eq:FP_LLE_t11}
    \nonumber   t_R \frac{\partial \psi}{\partial t} &= -(\alpha-i\phi_0)\psi -2iL \frac{\beta_2}{2\beta_1^2}\frac{\partial^2 \psi}{\partial z^2} +\theta_1 E_{in}(\beta_1 z)\\
    &+2iL\gamma \left( |\psi|^2 + \frac{G}{2L} \int_{-L}^L |\psi(z',t)|^2 dz' \right)\psi,
\end{align}
where $\alpha = \kappa t_R=-\ln(\rho_1 \rho_2) \approx 1 - \rho_1\rho_2$.
This form of FP-LLE reduces to the one obtained by Cole \textit{et al}.~\cite{cole2018theory} for the case of CW pumping and identical mirrors.
Its structure is usual in the context of microresonators~\cite{chembo2013spatiotemporal}.
More precisely, the evolution is in time and the transverse dimension is the space with periodic boundary conditions.

In fiber ring resonators it is customary to have evolution in space (also called slow time) and a temporal transverse coordinate \cite{haelterman1992additive,coen2013modeling}.
The role of time and space can be swapped at first order if we consider that the most important effect is the translation at the group velocity~\cite{chabchoub2016hydrodynamic}.
Indeed, in (\ref{eq:FP_LLE_t1}) the first two terms are of order one, while the remaining ones are first order corrections.
This means that, at the lowest order, we have
\begin{equation}
    \frac{\partial \psi}{\partial z} \approx -\beta_1 \frac{\partial \psi}{\partial t}\;, \quad \text{and} \quad \frac{\partial^2 \psi}{\partial z^2} \approx \beta_1^2 \frac{\partial \psi}{\partial t^2} 
\end{equation}
By using the second of the relations above in Eq. (\ref{eq:FP_LLE_t1}) and making the change of variable $t \rightarrow t-\beta_1z$, we get the space propagated version of the FP-LLE.
\begin{align}\label{eq:FP_LLE_t2}
   \nonumber  2L \frac{\partial \psi}{\partial z} &= -(\alpha-i\phi_0)\psi -iL\beta_2 \frac{\partial^2 \psi}{\partial t^2} + \theta_1 E_{in} \\
   &+2i\gamma L \left( |\psi|^2 + \frac{G}{t_R} \int_{-t_R/2}^{t_R/2} |\psi(z,t')|^2 dt' \right)\psi,
\end{align}
where $z>0$ and $-t_R/2 <t<t_R/2$.
Even if Eq.~(\ref{eq:FP_LLE_t2}) and Eq.~(\ref{eq:FP_LLE_t1}) have the same degree of approximation, only the time-propagated version has the correct boundary conditions.
Indeed, in Eq.~(\ref{eq:FP_LLE_t2}) we have assumed that the field is periodic in time, which is not strictly true.
This also implies that the modes have a constant frequency spacing (free spectral range, FSR), while in reality the FSR changes slightly because of dispersion.
Conversely, in Eq.~(\ref{eq:FP_LLE_t11}) the modes have equally spaced wavenumbers, but their frequencies are fixed by the dispersion relation.
These facts are almost irrelevant in standard (i.e. 'long', tens of meters) fiber ring resonators, because the roundtrip time is usually much longer than the pulse circulating in the resonator.
This usually allows one to consider an infinite roundtrip time with constant boundary conditions.
The field is no more considered as periodic and its spectrum, which is now continuous, gives the envelope of the discrete-spectrum of the full optical field circulating in the cavity.

\section{Exact solution for square pulse pumping}\label{sec:pulses_exact}

In this section we report the explicit expressions of the cross-phase modulation terms Eqs.~(\ref{eq:x_int}), for  $t\in[0,t_R]$, given the periodicity of the functions.
The expressions are different depending if the duty-cycle $f_r$ of the square pulse is greater or lesser than $0.5$.
The fundamental period $[0,t_R]$ is divided into six intervals, where the functions Eqs.~(\ref{eq:x_int}) have different forms.
For each time interval, there exist three different spatial intervals where the functions (\ref{eq:x_int}) are different in general.
Tables \ref{tab:table1} and \ref{tab:table2} report the explicit expressions of Eqs.~(\ref{eq:x_int}) for $f_r>0.5$ and $f_r<0.5$.
\begin{widetext}
    
\begin{table*}[htb]
    \label{tab:table2}
    \begin{tabular}{|c|c|c|c|c|}
    \hline
    \multicolumn{4}{|c|}{$f_r<0.5$}\\ \hline
    Time interval & Space interval & $\phi_{XF}(z,t)$ & $\phi_{XB}(z,t)$\\ \hline
    \multirow{2}{*}{$\displaystyle 0<t<\frac{t_R}{2}f_r$} & $\displaystyle 0<z<\frac{t}{\beta_1}$ & $|B_0|^2z $ &  $|F_0|^2z $\\
        & $\displaystyle\frac{t}{\beta_1}<z<2f_rL-\frac{t}{\beta_1}$ &0 &$\displaystyle|F_0|^2\left(\frac{z}{2}+\frac{t}{2\beta_1}\right)$\\
        & $\displaystyle 2f_rL-\frac{t}{\beta_1}<z<L$ &0 &$0$\\ \hline
    \multirow{3}{*}{$\displaystyle \frac{t_R}{2}f_r<t<t_Rf_r$} & $\displaystyle 0<z<2f_rL-\frac{t}{\beta_1}$ & $|B_0|^2z $ &  $|F_0|^2z $\\
        & $\displaystyle 2f_rL-\frac{t}{\beta_1}<z<\frac{t}{\beta_1}$ & $\displaystyle|B_0|^2\left(\frac{z}{2}-\frac{t}{2\beta_1}+f_rL\right)$ & $0$\\
        & $\displaystyle \frac{t}{\beta_1}<z<L$ & $0$ & $0$\\ \hline
    \multirow{3}{*}{$\displaystyle t_Rf_r<t<\frac{t_R}{2}$} & $\displaystyle 0<z<\frac{t}{\beta_1}-2f_rL$ & $0$ &  $0$\\
        & $\displaystyle \frac{t}{\beta_1}-2f_rL<z<\frac{t}{\beta_1}$ & $\displaystyle|B_0|^2\left(\frac{z}{2}-\frac{t}{2\beta_1}+f_rL\right)$ &$0$\\
        & $\displaystyle \frac{t}{\beta_1}<z<L$ & $0$ & $0$\\\hline
    \multirow{3}{*}{$\displaystyle \frac{t_R}{2}<t<\frac{t_R}{2}(1+f_r)$} & $\displaystyle 0<z<\frac{t}{\beta_1}-2f_rL$ & $0$ &  $0$\\
        & $\displaystyle \frac{t}{\beta_1}-2f_rL<z<2L-\frac{t}{\beta_1}$ & $\displaystyle|B_0|^2\left(\frac{z}{2}-\frac{t}{2\beta_1}+f_rL\right)$ &$0$\\
        & $\displaystyle 2L-\frac{t}{\beta_1}<z<L$ & $|B_0|^2(z-L(1-f_r))$ & $|F_0|^2(z-L(1-f_r))$\\\hline
    \multirow{3}{*}{$\displaystyle \frac{t_R}{2}(1+f_r)<t<t_R\left(\frac{1}{2}+f_r\right)$} & $\displaystyle 0<z<2L-\frac{t}{\beta_1}$ & $0$ &  $0$\\
        & $\displaystyle 2L-\frac{t}{\beta_1}<z<\frac{t}{\beta_1}-2f_rL$ & $0$ &$\displaystyle|F_0|^2\left(\frac{z}{2}+\frac{t}{2\beta_1}-L\right)$\\
        & $\displaystyle \frac{t}{\beta_1}-2f_rL<z<L$ & $|B_0|^2(z-L(1-f_r))$ & $|F_0|^2(z-L(1-f_r))$\\\hline
    \multirow{3}{*}{$\displaystyle t_R\left(\frac{1}{2}+f_r\right)<t<t_R$} & $\displaystyle 0<z<2L-\frac{t}{\beta_1}$ & $0$ &  $0$\\
        & $\displaystyle 2L- \frac{t}{\beta_1}<z<2L(1+f_r)-\frac{t}{\beta_1}$ & $0$ &$\displaystyle|F_0|^2\left(\frac{z}{2}+\frac{t}{2\beta_1}-L\right)$\\
        & $\displaystyle 2L(1+f_r)-\frac{t}{\beta_1}<z<L$ & $0$ & $0$\\\hline
    \end{tabular}
    \caption{Cross-phase terms for $f_r<0.5$.}
\end{table*}

\begin{table*}
    \label{tab:table1}
    \begin{tabular}{|c|c|c|c|c|}
        \hline
        \multicolumn{4}{|c|}{$f_r>0.5$}\\ \hline
        Time interval & Space interval & $\phi_{XF}(z,t)$ & $\phi_{XB}(z,t)$\\ \hline
        \multirow{3}{*}{$\displaystyle 0<t<t_R\left(f_r-\frac{1}{2}\right)$}  & $\displaystyle 0<z<\frac{t}{\beta_1}$ & $|B_0|^2z $ & $|F_0|^2z $\\
            & $\displaystyle\frac{t}{\beta_1}<z<\frac{t}{\beta_1}+2L(1-f_r)$ &0 &$\displaystyle|F_0|^2\left(\frac{z}{2}+\frac{t}{2\beta_1}\right)$\\
            & $\displaystyle\frac{t}{\beta_1}+2L(1-f_r)<z<L$ & $|B_0|^2(z-L(1-f_r)) $ & $|F_0|^2(z-L(1-f_r))$\\ \hline
        \multirow{3}{*}{$\displaystyle t_R\left(f_r-\frac{1}{2}\right)<t<\frac{t_R}{2}f_r$} & $\displaystyle 0<z<\frac{t}{\beta_1}$ & $|B_0|^2z $ &  $|F_0|^2z $\\
            & $\displaystyle\frac{t}{\beta_1}<z<2f_rL-\frac{t}{\beta_1}$ &0 &$\displaystyle|F_0|^2\left(\frac{z}{2}+\frac{t}{2\beta_1}\right)$\\
            & $\displaystyle 2f_rL-\frac{t}{\beta_1}<z<L$ & $0$ & $0$\\ \hline
        \multirow{3}{*}{$\displaystyle \frac{t_R}{2}f_r<t<\frac{t_R}{2}$} & $\displaystyle 0<z<2f_rL-\frac{t}{\beta_1}$ & $|B_0|^2z $ &  $|F_0|^2z $\\
            & $\displaystyle 2f_rL-\frac{t}{\beta_1}<z<\frac{t}{\beta_1}$ & $\displaystyle|B_0|^2\left(\frac{z}{2}-\frac{t}{2\beta_1}+f_rL\right)$ &$0$\\
            & $\displaystyle \frac{t}{\beta_1}<z<L$ & $0$ & $0$\\ \hline
        \multirow{3}{*}{$\displaystyle \frac{t_R}{2}<t<t_Rf_r$} & $\displaystyle 0<z<2f_rL-\frac{t}{\beta_1}$ & $|B_0|^2z $ &  $|F_0|^2z $\\
            & $\displaystyle 2f_rL-\frac{t}{\beta_1}<z<2L-\frac{t}{\beta_1}$ & $\displaystyle|B_0|^2\left(\frac{z}{2}-\frac{t}{2\beta_1}+f_rL\right)$ &$0$\\
            & $\displaystyle 2f_rL-\frac{t}{\beta_1}<z<L$ & $|B_0|^2(z-L(1-f_r))$ & $|F_0|^2(z-L(1-f_r))$\\ \hline
        \multirow{3}{*}{$\displaystyle t_Rf_r<t<\frac{t_R}{2}(1+f_r)$} & $\displaystyle 0<z<\frac{t}{\beta_1}-2f_rL$ & $0$ &  $0$\\
            & $\displaystyle \frac{t}{\beta_1}-2f_rL<z<2L-\frac{t}{\beta_1}$ & $\displaystyle|B_0|^2\left(\frac{z}{2}-\frac{t}{2\beta_1}+f_rL\right)$ &$0$\\
            & $\displaystyle 2L-\frac{t}{\beta_1}<z<L$ & $|B_0|^2(z-L(1-f_r))$ & $|F_0|^2(z-L(1-f_r))$\\ \hline
        \multirow{3}{*}{$\displaystyle \frac{t_R}{2}(1+f_r)<t<t_R$} & $\displaystyle 0<z<2L-\frac{t}{\beta_1}$ & $0$ &  $0$\\
            & $\displaystyle 2L- \frac{t}{\beta_1}<z<\frac{t}{\beta_1}-2f_rL$ & $0$ &$\displaystyle|F_0|^2\left(\frac{z}{2}+\frac{t}{2\beta_1}-L\right)$\\
            & $\displaystyle \frac{t}{\beta_1}-2f_rL<z<L$ & $|B_0|^2(z-L(1-f_r))$ & $|F_0|^2(z-L(1-f_r))$\\ \hline
    \end{tabular}
    \caption{Cross-phase terms for $f_r>0.5$.}
\end{table*}

\end{widetext}

\providecommand{\noopsort}[1]{}\providecommand{\singleletter}[1]{#1}%

\end{document}